\documentclass[fleqn,usenatbib]{mnras}
\usepackage{float}
\usepackage{newtxtext,newtxmath}
\usepackage{afterpage}
\usepackage{placeins}
\usepackage[T1]{fontenc}
\usepackage{xcolor}
\DeclareRobustCommand{\VAN}[3]{#2}
\let\VANthebibliography\thebibliography
\def\thebibliography{\DeclareRobustCommand{\VAN}[3]{##3}\VANthebibliography}
\usepackage{longtable}
\usepackage{supertabular}

\usepackage{orcidlink}
\usepackage{graphicx}	% Including figure files
\usepackage{amsmath}	% Advanced maths commands

\usepackage{orcidlink}

\title[Timing properties of RX J0440.9+4431]{Timing properties of the X-ray accreting pulsar RX J0440.9+4431 studied with Insight-HXMT and NICER}

\author[P. P. Li et al.]{
P. P. Li,$^{1,2}$ L. Tao~\orcidlink{0000-0002-2705-4338},$^{1}$\thanks{E-mail: taolian@ihep.ac.cn} Y. L. Tuo~\orcidlink{0000-0003-3127-0110},$^{3}$ M. Y. Ge~\orcidlink{0000-0002-2749-6638},$^{1}$ L. D. Kong~\orcidlink{0000-0003-3188-9079},$^{3}$ L. Zhang,$^{1}$ Q. C. Bu~\orcidlink{0000-0001-5238-3988},$^{3}$ L. Ji~\orcidlink{0000-0001-9599-7285},$^{4}$ \newauthor{J. L. Qu,$^{1}$ S. Zhang,$^{1}$ S. N. Zhang~\orcidlink{0000-0001-5586-1017},$^{1}$ Y. Huang,$^{1}$ X. Ma,$^{1}$  W. T. Ye,$^{1, 2}$ Q. C. Zhao,$^{1, 2}$ } R. C. Ma,$^{1, 2}$ \newauthor{S. J. Zhao,$^{1, 2}$ X. Hou,$^{5,6}$ Z. X. Yang,$^{7}$ P. J. Wang,$^{1,2}$ S. M. Jia,$^{1}$ Q. C. Shui,$^{1, 2}$ J. Guan,$^{1}$}
\\
$^{1}$Key Laboratory of Particle Astrophysics, Institute of High Energy Physics, Chinese Academy of Sciences, 100049 Beijing, People’s Republic of China\\
$^{2}$Uinversity of Chinese Academy of Sciences, Chinese Academy of Sciences, 100049 Beijing, People’s Republic of China\\
$^{3}$Institut f\"ur Astronomie und Astrophysik, Kepler Center for Astro and Particle Physics, Eberhard Karls Universit\"at, Sand 1, D-72076 T\"ubingen, Germany\\
$^{4}$School of Physics and Astronomy, Sun Yat-sen University, Zhuhai, 519082, People’s Republic of China\\
$^{5}$Yunnan Observatories, Chinese Academy of Sciences, Kunming 650216, People's Republic of China\\
$^{6}$Key Laboratory for the Structure and Evolution of Celestial Objects, Chinese Academy of Sciences, Kunming 650216, People’s Republic of China\\ 
$^{7}$School of Physics and Optoelectronic Engineering, Shandong University of Technology, Zibo 255000, People’s Republic of China}

\date{Accepted XXX. Received YYY; in original form ZZZ}

\pubyear{2015}

\begin{document}
\label{firstpage}
\pagerange{\pageref{firstpage}--\pageref{lastpage}}
\maketitle

\begin{abstract}
RX J0440.9+4431, a Be/X-ray binary, had its brightest outburst in 2022 since its discovery, with a peak X-ray flux of 2.25 Crab (as recorded by \textit{Swift}/BAT, 15--50 keV). We analyze the timing properties of this giant outburst using data from \textit{Insight}-HXMT and \textit{NICER}, focusing on the evolution of the pulse profile and pulse fraction. We observe that when the luminosity reached around \textasciitilde$3\times10^{37}\ {\rm erg\ \rm s^{-1}}$, a transition from double-peaked to single-peaked pulse profiles occurred across the energy range, with the peak of the low-energy profile aligning gradually with the peak of the high-energy profile. This change indicates a transition from subcritical to supercritical accretion. Additionally, we found a concave in the pulse fraction as a function of energy around 20--30\,keV throughout the entire outburst period. Compared to the low luminosity, the concave becomes weaker in high luminosities, and overall, the pulse fraction is higher. We propose that this concave could be caused by the scattering of high-energy photons by the atmosphere of a neutron star, leading to a dilution of the pulse fraction. As the accretion reaches the supercritical state, the accretion column height increases, resulting in a larger direct component of strongly beamed X-ray flux, and an elevated pulse fraction.

\end{abstract}


%%%%%%%%%%%%%%%%%%%%%%%%%%%%%%%%%%%%%%%%%%%%%%%%%%

%%%%%%%%%%%%%%%%% BODY OF PAPER %%%%%%%%%%%%%%%%%%

\section{Introduction}
Be/X-ray binaries (BeXRBs) are a subclass of high-mass X-ray binaries (HMXBs). Generally, they consist of a Be star as the companion and a magnetized ($\geq 10^{12}\ \text{G}$) neutron star (see reviews by \cite{2011Reig,2022Mushtukov}). Be stars are non-supergiant fast-rotating B-type and luminosity class III-V stars. As these stars rotate, they expel matter, giving rise to a disk formation encircling their equator, characterized by emission lines and an excess of infrared radiation \citep{2003Porter}. Generally, BeXRBs are eccentric systems ($e \geq 0.3$) with relatively wide orbits ($\geq 10\,d$). During periastron, the neutron star approaches the circumstellar disk or, regardless of orbital phase, crosses the circumstellar disk, causing significant disruption. Subsequently, a large amount of material is accumulated onto the neutron star. At the magnetospheric radius \citep{2015Belenkaya}, the material is captured by the magnetic field, and the kinetic energy of the falling matter along the magnetic field lines is transformed into radiation, providing energy for X-ray emission. Misalignment of rotational and magnetic axis results in the phenomenon of X-ray pulsations.

The geometry of the emission region in X-ray pulsars is defined by the magnetic field structure and mass accretion rate. At low luminosities (\textasciitilde$ 10^{34-35}\ {\rm erg\ \rm s^{-1}}$), the material passes through a conventional gas-mediated shock, forming a hot spot on the surface of the neutron star, and the emitted radiation escapes from the top of column, forming a "pencil beam" \citep{1993Nelson}. At high luminosities (\textasciitilde$ 10^{37-38}\ {\rm erg\ \rm s^{-1}}$), the material encounters a radiation-dominated shock near the top of the accretion column, ultimately settling onto the surface of the neutron star, and the emitted radiation escapes from the walls of the column, forming a "fan beam" \citep{1976Basko}. In the intermediate range, the emission pattern may be a mixture of these two types \citep{2000Blum,2012Becker}. Naturally, changes in the emission pattern during outbursts lead to variations in the observed pulse profiles. The pulse profiles depend on both the energy band and the accretion luminosity, reflecting the dependence of the beam pattern on the energy and mass accretion rate (see e.g. \cite{2010Tsygankov,2020wang0243,2022Wang0535,2021Rai,2022Hou}). Therefore, investigating pulse profiles provides valuable insights into the understanding of X-ray pulsar accretion mechanisms.

%RX J0440.9+4431 was initially discovered as a Be/X-ray binary in the \textit{ROSAT} Galactic plane survey data, and its optical companion was identified as LS V +44 17/BSD 24-491 \citep{1997Motch}. The distance of the source was most recently estimated to be 2.4\,kpc \citep{2021Bailer-Jones}. The \textit{RXTE}/PCA was the first to detect X-ray pulsations from RX J0440.9+4431, with a pulse period of $202.5\pm{0.5}$\,s and a sinusoidal pulse profile \citep{1999Reig}. \cite{2012Palombara} also found a single sinusoidal pulse profile in \textit{XMM-Newton} data, which was recorded in the RX J0440.9+4431 quiescent states on March 18, 2011. However, the pulse period of $204.96\pm{0.02}$\,s was larger than the previous measurement by \cite{1999Reig}, suggesting an average pulsar spin-down.

RX J0440.9+4431 was initially discovered as a Be/X-ray binary in the \textit{ROSAT} Galactic plane survey data, and its optical companion was identified as LS V +44 17/BSD 24-491 \citep{1997Motch}. The distance of the source was most recently estimated to be 2.4\,kpc \citep{2021Bailer-Jones}. The \textit{RXTE}/PCA was the first to detect X-ray pulsations from RX J0440.9+4431, with a pulse period of $202.5\pm{0.5}$\,s \citep{1999Reig}. However, subsequent analysis by \cite{2012Palombara} using \textit{XMM-Newton} data recorded during the RX J0440.9+4431 quiescent states on March 18, 2011, revealed a pulse period of $204.96\pm{0.02}$\,s, indicating an average pulsar spin-down.

%However, \cite{2012Palombara} found the pulse period of $204.96\pm{0.02}$\,s in \textit{XMM-Newton} data, which was recorded in the RX J0440.9+4431 quiescent states on March 18, 2011, suggesting an average pulsar spin-down.

%As Fig.~\ref{fig:1} shows, 
RX J0440.9+4431 has experienced three consecutive flux increases until 2022, none of which have a luminosity greater than $10^{37}\ {\rm erg\ \rm s^{-1}}$. In late March of 2010, the first recorded outburst from RX J0440.9+4431 was detected using the Monitor of All-sky X-ray Image (\textit{MAXI}) instrument \citep{2010ATelMorii}. Then, \cite{2012Usui} analyzed three observations using \textit{RXTE}/PCA and discovered an absorption dip structure at a pulse phase of  \textasciitilde0.25 after the main peak in the lower energy bands during an observation that was near the most luminous period of the first outburst. The absorption column density at the dip phase was much higher than those in the other phases. Based on this, \cite{2012Usui} believes that the dip in the pulse profile is caused by the eclipse of radiation from the neutron star by the accretion column. The second outburst from RX J0440.9+4431 was detected by \textit{INTEGRAL} on September 1, 2010, with a luminosity less than half of the previous outburst. According to the analysis conducted by \cite{2012Tsygankov} using data from \textit{Swift}/XRT and \textit{RXTE}/PCA, the pulse profiles exhibit a simple sine-like shape. Additionally, the pulse fraction decreases with decreasing energy in the range of 4--20\,keV. Subsequently, in January 2011, the \textit{Swift} observatory detected the third outburst. The recurrence intervals of the three outbursts are approximately 150 days, leading \cite{2012Tsygankov} and \cite{2013Ferrigno} to suggest that the interval may be an orbital period of the binary star, assuming that these outbursts are Type I outbursts that occurred at the periastron passage of the neutron star. However, to accurately determine the orbital parameters of the binary system, much longer observations with high sensitivity and timing accuracy are required.
%The analysis conducted by \cite{2012Tsygankov} on the data from \textit{Swift}/XRT and \textit{RXTE}/PCA below 10\,keV did not reveal a dip in the pulse profile, which remained a sinusoid.

Eleven years after the third outburst, \textit{MAXI}/GSC detected a brightening of an X-ray source located at RX J0440.9+4431 \citep{2022Nakajima}. Interestingly, towards the end of this outbreak, its count rate increased once again \citep{2023ATelpal}, reaching a peak X-ray flux of 2.25 Crab \citep{2023ATelColey}, as recorded by \textit{Swift}/BAT (15--50\,keV). This unprecedented event allows for the study of the accretion process of RX J0440.9+4431 across a wider range of luminosity with high-quality data. In this paper, we analyze the dense observations of this source by \textit{Insight}-HXMT and \textit{NICER} to characterize its temporal properties, with a focus on the evolution of pulse profile that is energy and luminosity dependent. From the high-energy data that benefited from \textit{Insight}-HXMT, we can further understand the emission beam pattern of accretion columns in different accretion states. Section~\ref{sec:2} presents the observation and data reduction procedures, while Section~\ref{sec:3} provides a summary of the results obtained. Finally, Section~\ref{sec:4} discusses the findings.

%\begin{figure}
%	% To include a figure from a file named example.*
%	\includegraphics[width=\columnwidth]{1batlc.eps}
 %   \caption{The X-ray light curve of RX J0440.9+4431 obtained with the \textit{Swift}/BAT monitor in the 15--50\,keV energy band. The arrows indicate the locations of the three outbursts before this outburst in 2022-2023, respectively.}
  %\label{fig:1}
%\end{figure}

\section{Observations and Data Reduction}
\label{sec:2}
\subsection{\textit{Insight}-HXMT}
\label{sec:2.1}
The Hard X-ray Modulation Telescope (HXMT), also known as \textit{Insight}-HXMT, is the Chinese first X-ray astronomy satellite with a broad energy in 1--250\,keV \citep{zhang2020}. It has three main payloads, which are Low Energy X-ray telescope (LE; 1--15\,keV), Medium Energy X-ray telescope (ME; 5--30\,keV), and High Energy X-ray telescope (HE; 20--250\,keV). Starting from December 31, 2022, shortly after the MAXI trigger, \textit{Insight}-HXMT commenced its monitoring of RX J0440.9+4431. These observations continued until March 26, 2023, and ObsIDs used in this paper are listed in Table~\ref{tab:hxmt}.

All observations are analyzed using the \textit{Insight}-HXMT processing software {\tt HXMTDAS v2.05}\footnote{\url{http://hxmtweb.ihep.ac.cn/software.jhtml}} to generate clean and calibrated event data, as well as high-level data products such as spectra and light curves\footnote{\url{http://hxmtweb.ihep.ac.cn/SoftDoc/648.jhtml}}. The clean event file of the source is obtained with the following criteria: (1) the Sun Angle $>70^{\circ}$; (2) the Moon Angle $>6^{\circ}$; (3) the elevation angle $>10^{\circ}$; (4) the Cut-off Rigidity value $>8$\,GV; (5) the pointing offset angle $<0.04^{\circ}$. In addition, the arrival time of photons in the clean event file is corrected using the {\tt hxbary} tool.

%#####################################################################
\begin{figure}
	\includegraphics[width=0.95\columnwidth]{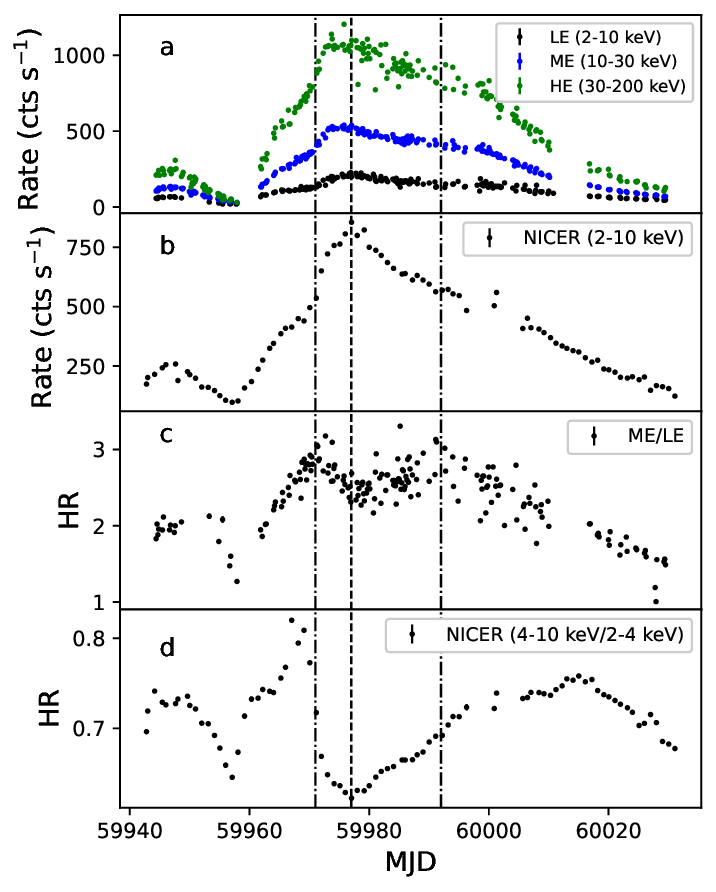}
    \caption{Panel a: X-ray light curve of RX J0440.9+4431 by LE (black), ME (blue), and HE (green) with each data binned by one sub-exposure ID (ExpID). Panel b: X-ray light curve of RX J0440.9+4431 in the energy band of 2--10\,keV by \textit{NICER} with each data binned by one ObsID. Panel c: Hardness radio (HR) between 10--30\,keV (ME) and 2--10\,keV (LE) energy bands using \textit{Insight}-HXMT. Panel d: Hardness radio between 4--10\,keV and 2--4\,keV energy bands by \textit{NICER}. Two dashed dotted lines (MJD 59971; MJD 59992) indicate the transition of the pulse profile as shown in Fig.~\ref{fig:3}. The middle dashed line (MJD 59977) marks the peak of the X-ray light curve for RX J0440.9+4431.}
  \label{fig:1}
\end{figure}
%#####################################################################
\begin{figure}
	\includegraphics[width=0.9\columnwidth]{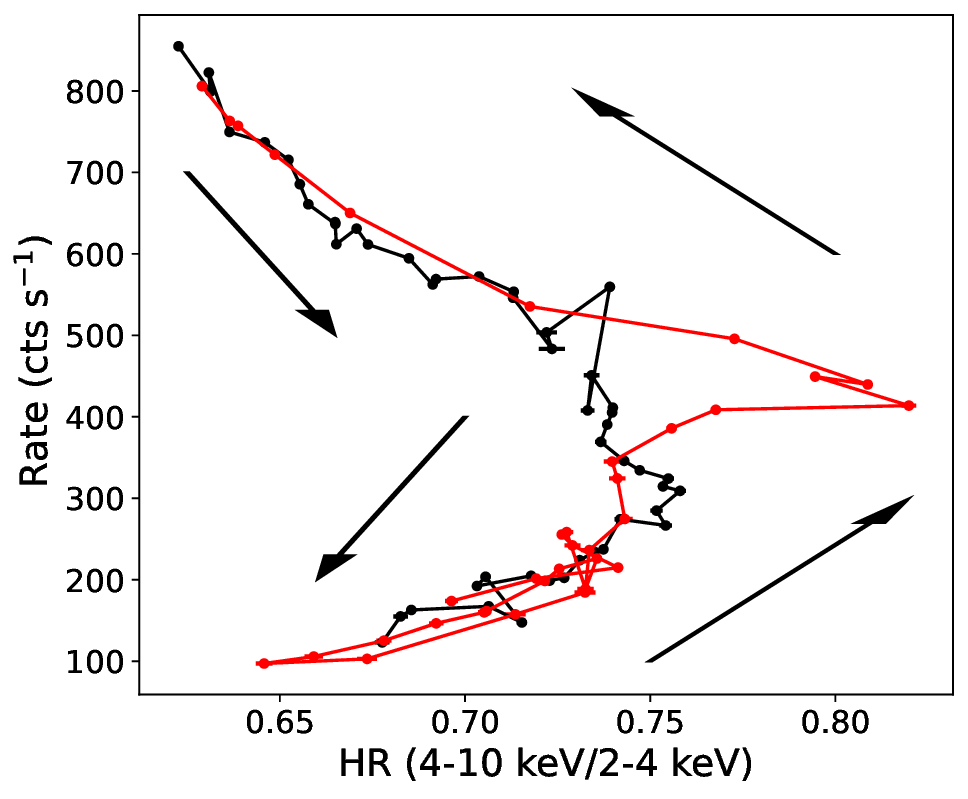}
    \caption{The hardness-intensity diagram (HID) of RX J0440.9+4431 in 2022--2023 giant outburst, where the hardness is defined as the count rate ratio of 4--10 to 2--4\,keV observed with \textit{NICER}. The black arrows describe the direction of evolution. The red and black colors represent the rising and declining parts of the outburst, respectively.}
  \label{fig:2}
\end{figure}
%#####################################################################

\subsection{\textit{NICER}}
\label{sec:2.2}
The Neutron star Interior Composition Explorer (\textit{NICER}) is an instrument onboard the International Space Station (ISS) to study neutron stars in the 0.2--12\,keV X-ray band \citep{2016SPIE}. \textit{NICER} also conducted extensive observations of RX J0440.9+4431 over a period of more than three months, starting on December 29, 2022 and ending on April 15, 2023. This paper uses observations taken before and on March 26, 2023, which are listed in Table~\ref{tab:nicer}.

To process these data, we utilize {\tt HEASoft v6.31.1} and the \textit{NICER} Data Analysis Software ({\tt NICERDAS 2022-12-16\_v010a}) with Calibration Database ({\tt CALDB vXTI20221001}). The {\tt nicerl2} tool is used to implement all standard calibration and screening criteria, producing clean event files. The {\tt barycorr} task is then applied to each clean event file to implement barycentric corrections. Finally, we use a single pipeline task called {\tt nicerl3-lc} to extract all light curve products. In order to reduce the impact of background, we examined the light curves of each observation in the 13-15\,keV range and removed the times with count rates greater than 1 counts s$^{-1}$ \citep{2020Zhangl}.

\section{Analysis and Results}
\label{sec:3}
\subsection{Light Curve, Hardness Ratio, and HIDs}
\label{sec:3.1}
Fig.~\ref{fig:1} describes the X-ray light curve and hardness ratio (HR) of RX J0440.9+4431 obtained by \textit{Insight}-HXMT and \textit{NICER}. In Panel a, the light curves for three energy bands -- LE (2--10\,keV), ME (10--30\,keV), and HE (30--200\,keV) -- obtained from \textit{Insight}-HXMT are displayed. Panel b shows the light curve from \textit{NICER} in the energy range of 2--10\,keV. The observations of RX J0440.9+4431 by both satellites begin when the count rate of the minor outburst is approaching its peak, with the minor outburst ending on MJD 59958. Subsequently, the count rate rise, and a more extreme outburst ensue. The middle dashed line in Fig.~\ref{fig:1} indicates the time of the highest count rate, which occurred on MJD 59977.
%As shown in panel b of Fig.~\ref{fig:1}, the maximum count rate of the subsequent giant outburst is more than twice as high as that of the minor outburst. 

In panel c of Fig.~\ref{fig:1}, we define the HR as the ratio of 10--30\,keV to 2--10\,keV energy bands in \textit{Insight}-HXMT. During the minor outburst, the HR increases as the count rate increases and decreases as the count rate decreases. As the source brightens again, the HR also increases once more. However, at MJD 59971, the hardness ratio begins to decrease. Then, after reaching the minimum around the time of the highest count rate of RX J0440.9+4431, the HR gradually increases again. Finally, HR decreases with a decrease in the count rate at MJD 59992. The panel d is the HR we get using the ratio of 4--10\,keV to 2--4\,keV of \textit{NICER}, and similar to \textit{Insight}-HXMT, HR decreases when it is MJD 59971, and recovers after the HR reaches its lowest when the source is the brightest. The HR and count rate of \textit{NICER} are plotted in the hardness-intensity diagram (HID) in Fig.~\ref{fig:2}, with arrows indicating the direction of evolution. The turning in Fig.~\ref{fig:2} corresponds to when HR in the panel d in Fig.~\ref{fig:1} sharply reduces (\textasciitilde MJD 59971), which aligns with the results presented in Figure 8 by \cite{2023Salganik}. This transition may be related to a change in the accretion state of the pulsar, similar to what was found in the Swift J0243.6+6124 \citep{2020Kong} and RX J0209.6--7427 \citep{2022Hou}.

%In panel c of Fig.~\ref{fig:1}, we define the HR as the ratio of 10--30\,keV to 2--10\,keV energy bands in \textit{Insight}-HXMT. During the minor outburst of RX J0440.9+4431, the HR increases as the count rate increases and decreases as the count rate decreases. As the source brightens again, the HR also increases once more. However, at MJD 59971, the hardness ratio begins to decrease, corresponding to the first dashed line in Fig.~\ref{fig:1}. Then, after reaching the minimum around the time of the highest count rate of RX J0440.9+4431, the HR gradually increases again. Finally, HR decreases with a decrease in the count rate at MJD 59992 (the third dashed line). The panel d is the HR we get using the ratio of 4--10\,keV to 2--4\,keV of \textit{NICER}, and like \textit{Insight}-HXMT, HR decreases when it is MJD 59971, and recovers after the HR reaches its lowest when the source is the brightest. The HR and count rate of \textit{NICER} are plotted in the hardness-intensity diagram (HID) in Fig.~\ref{fig:2}, with arrows indicating the direction of evolution. The turning in Fig.~\ref{fig:2} corresponds to when HR in the panel d in Fig.~\ref{fig:1} sharply reduces (\textasciitilde MJD 59971), and this transition may be related to a change in the accretion state of the pulsar, similar to what was found in the Swift J0243.6+6124 \citep{2020Kong} and  RX J0209.6--7427 \citep{2022Hou}.

%But the rate of recovery is much slower than that oif \textit{Insight}-HXMT.
%#####################################################################
\begin{figure*}
	\includegraphics[width=0.297\textwidth]{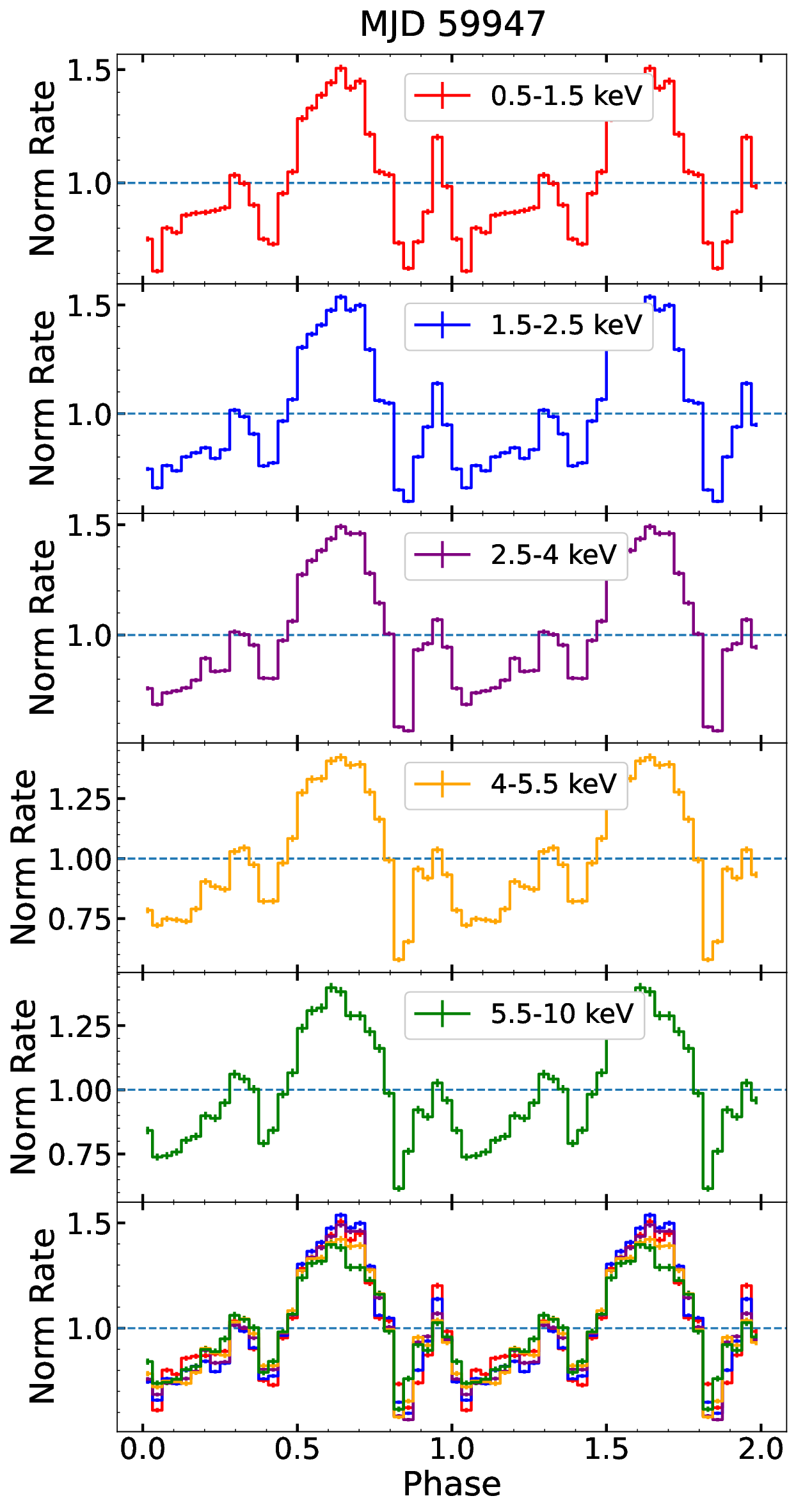}
    \includegraphics[width=0.29\textwidth]{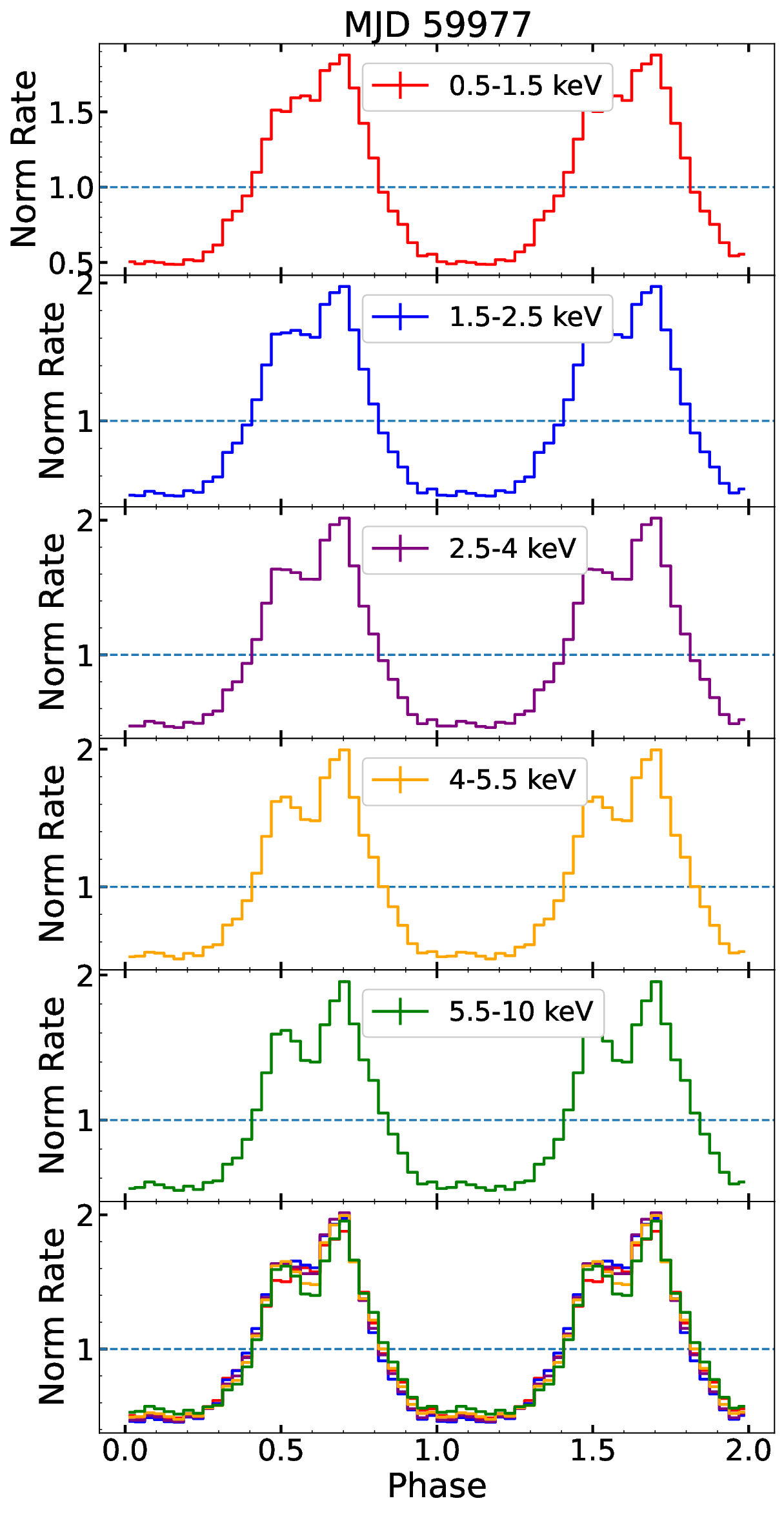}
    \includegraphics[width=0.29\textwidth]{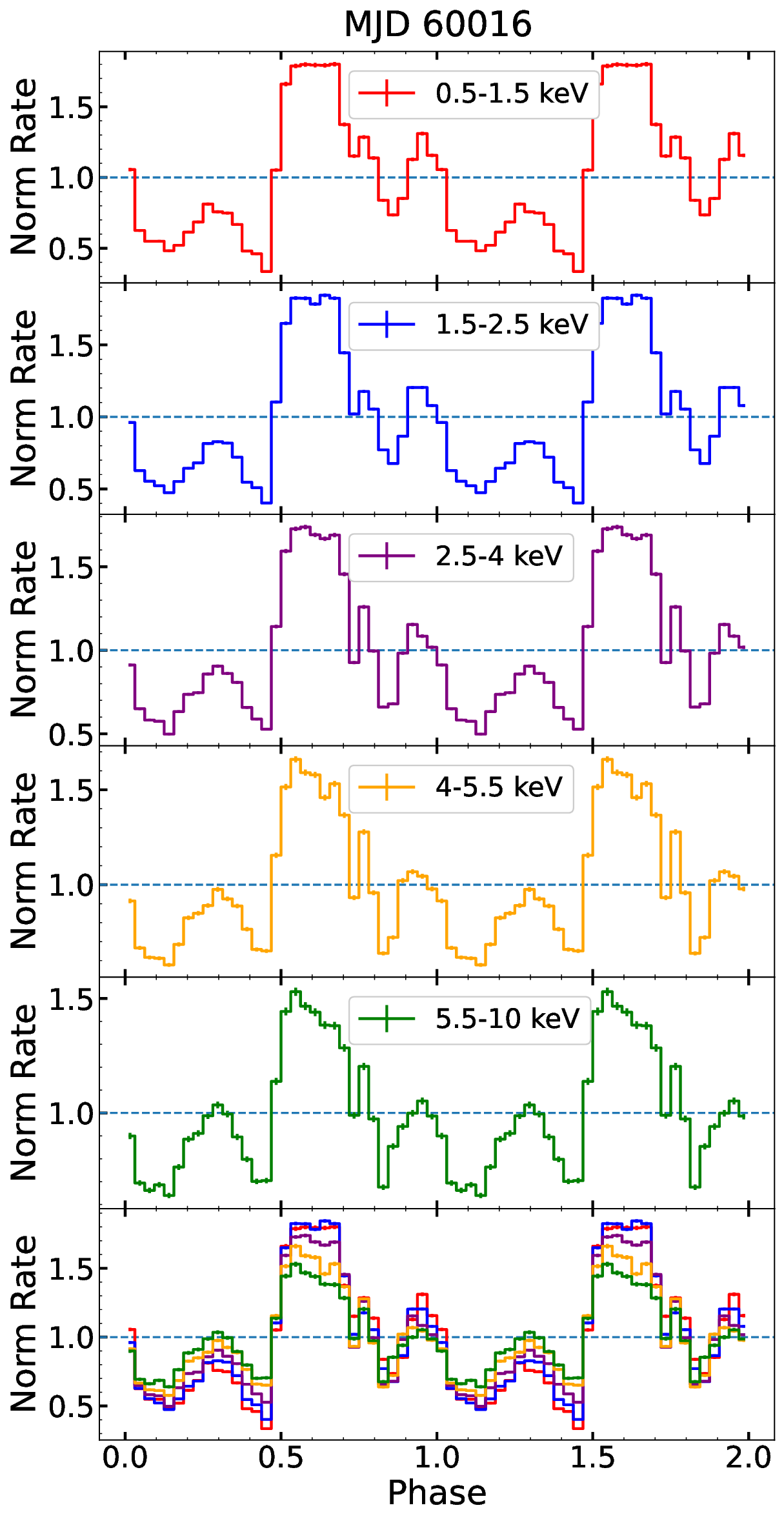}
    \includegraphics[width=0.298\textwidth]{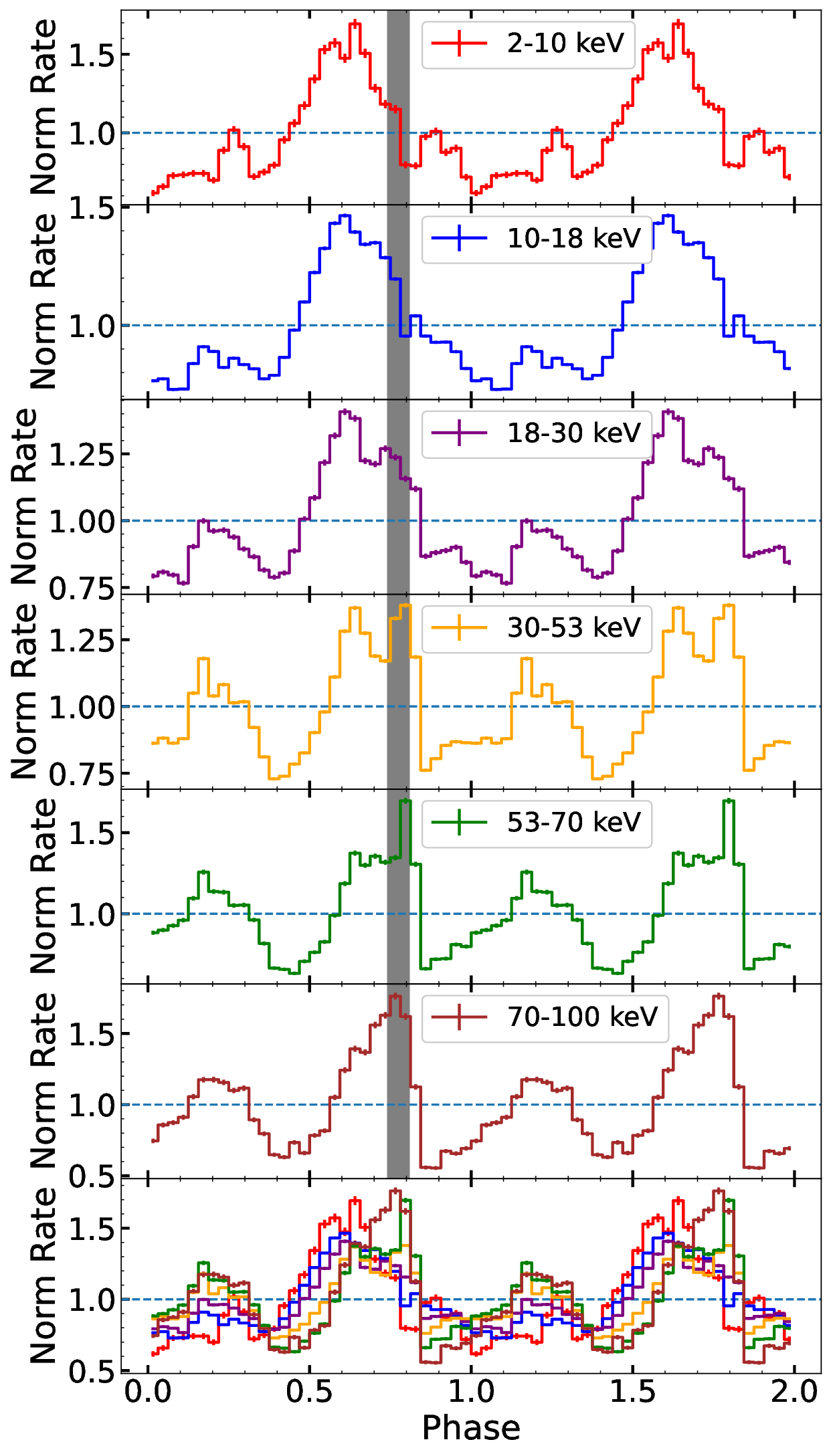}
    \includegraphics[width=0.28\textwidth]{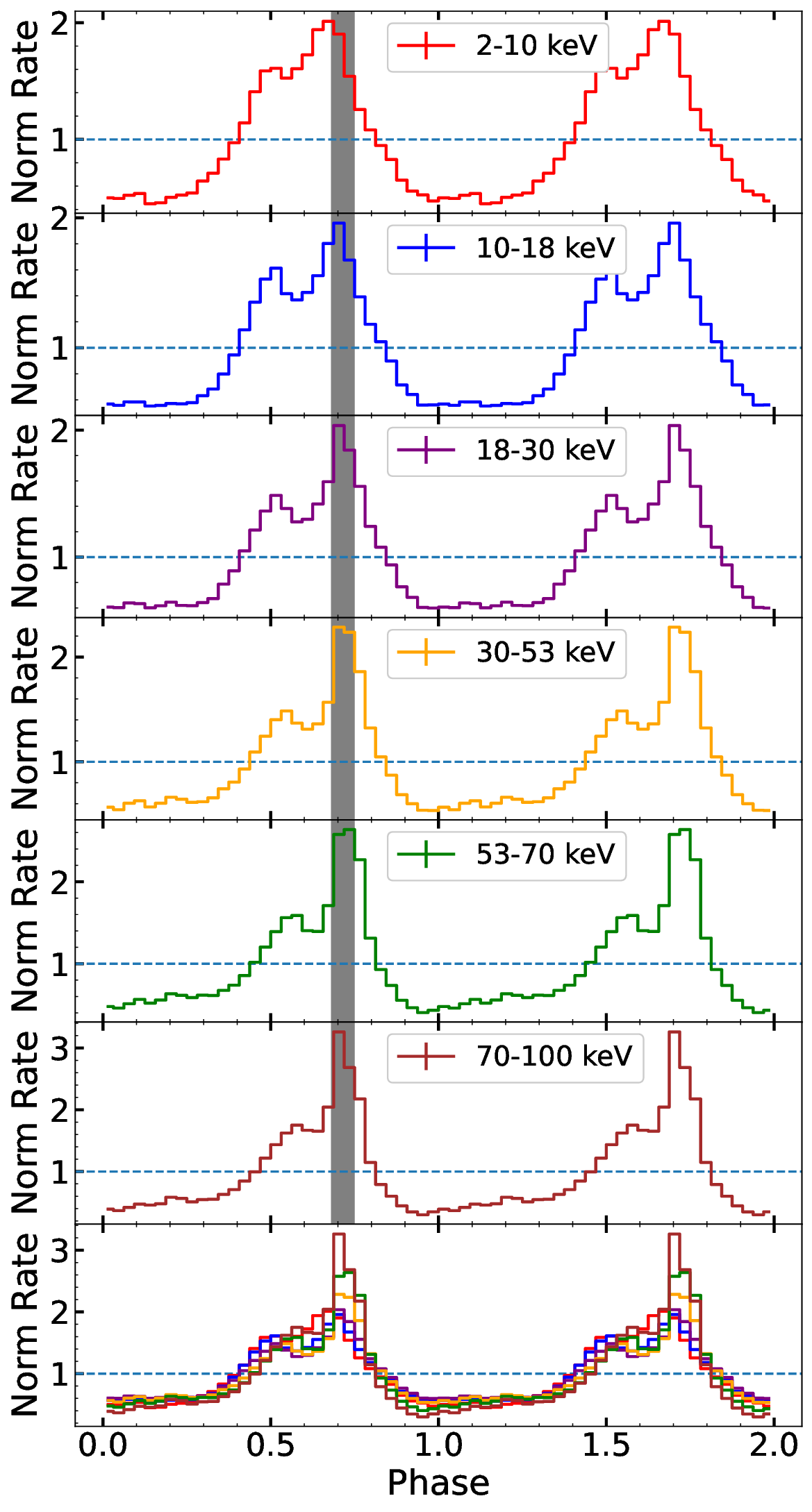}
    \includegraphics[width=0.298\textwidth]{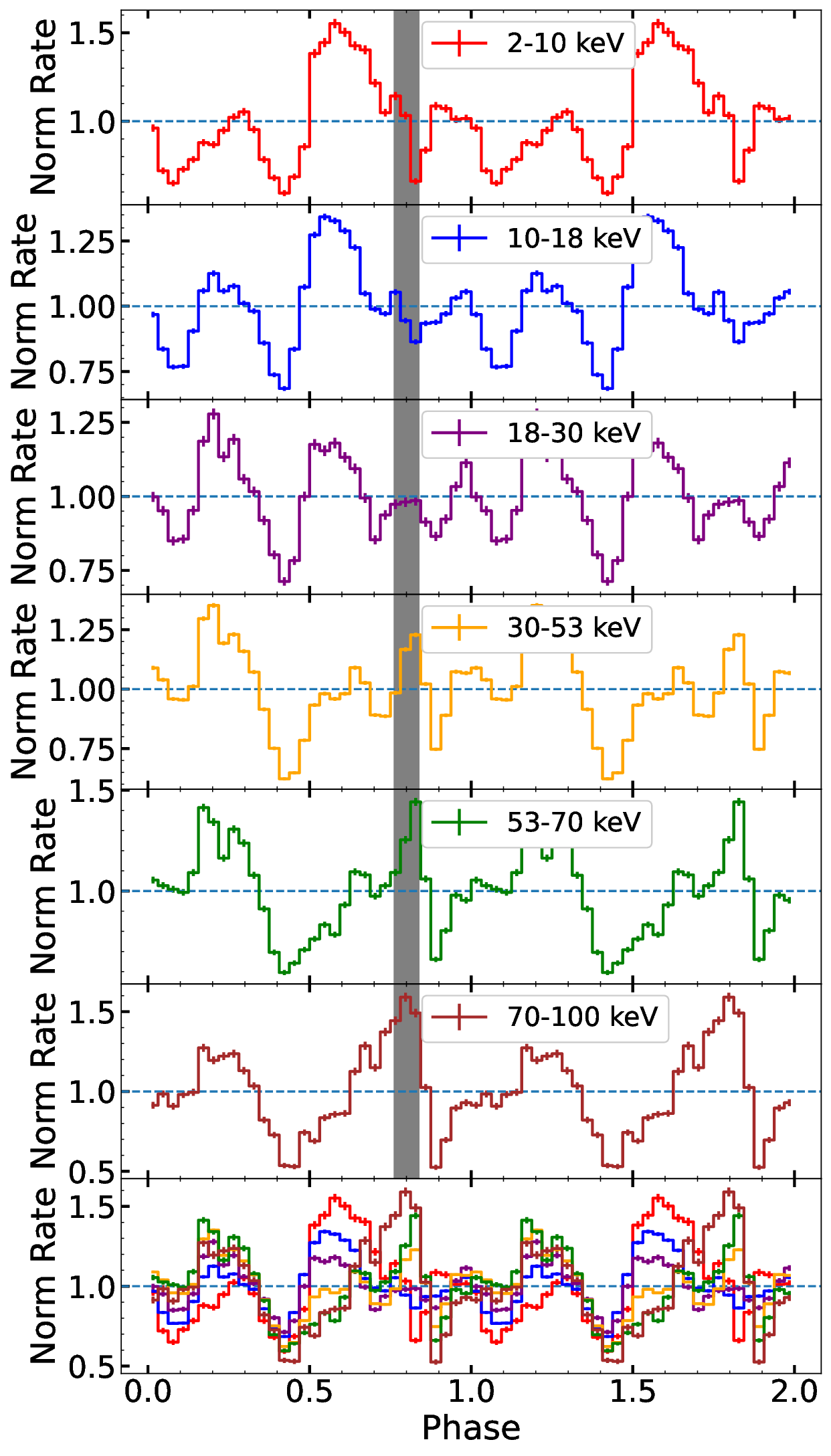}
    \caption{Variation of RX J0440.9+4431 pulse profiles with energy based on observations from \textit{NICER} (top row) and \textit{Insight}-HXMT (ME and HE; bottom row). Each pulse profile is normalized by its mean flux. The three columns from left to right represent pulse profiles from data taken on MJD 59947, MJD 59977, and MJD 60016, respectively. The pulse profiles for different energy ranges are re-plotted in the last row of each panel. The gray shading represents the phase of the peak where the high-energy (70--100\,keV) pulse profile is located.}
    
  \label{fig:4}
\end{figure*}

%#####################################################################

\begin{figure}
	\includegraphics[width=\columnwidth]{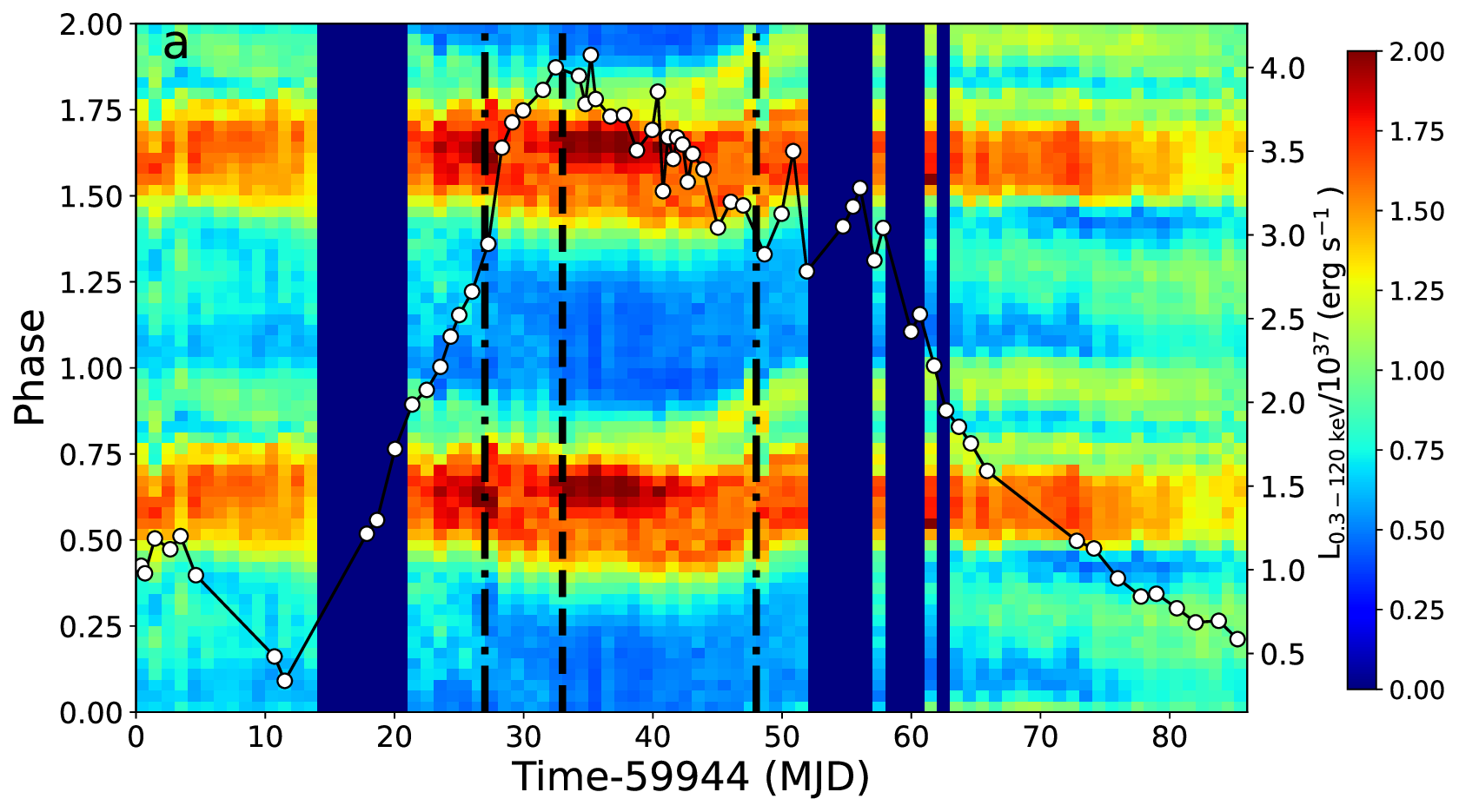}    \includegraphics[width=\columnwidth]{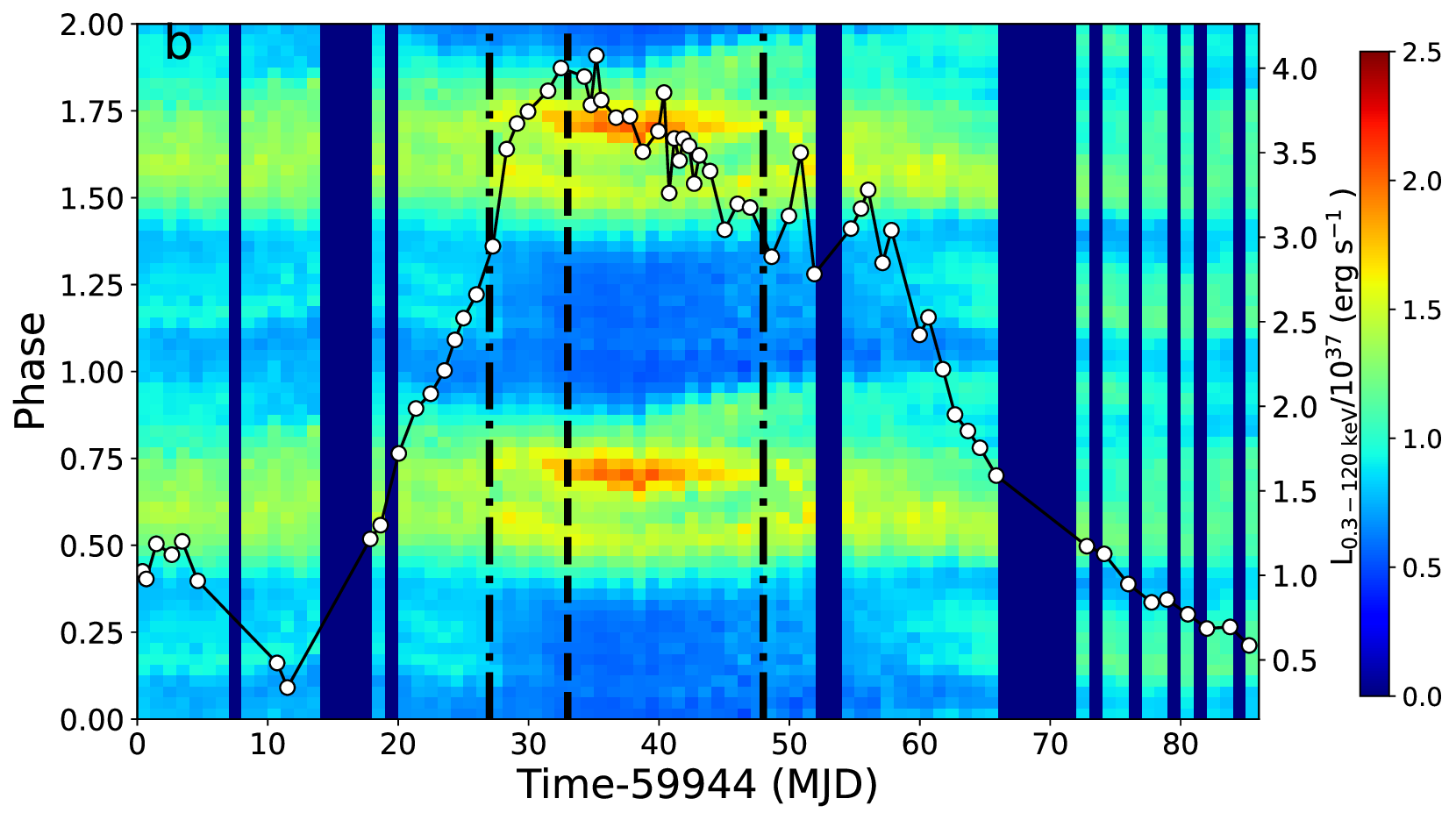}
    \includegraphics[width=\columnwidth]{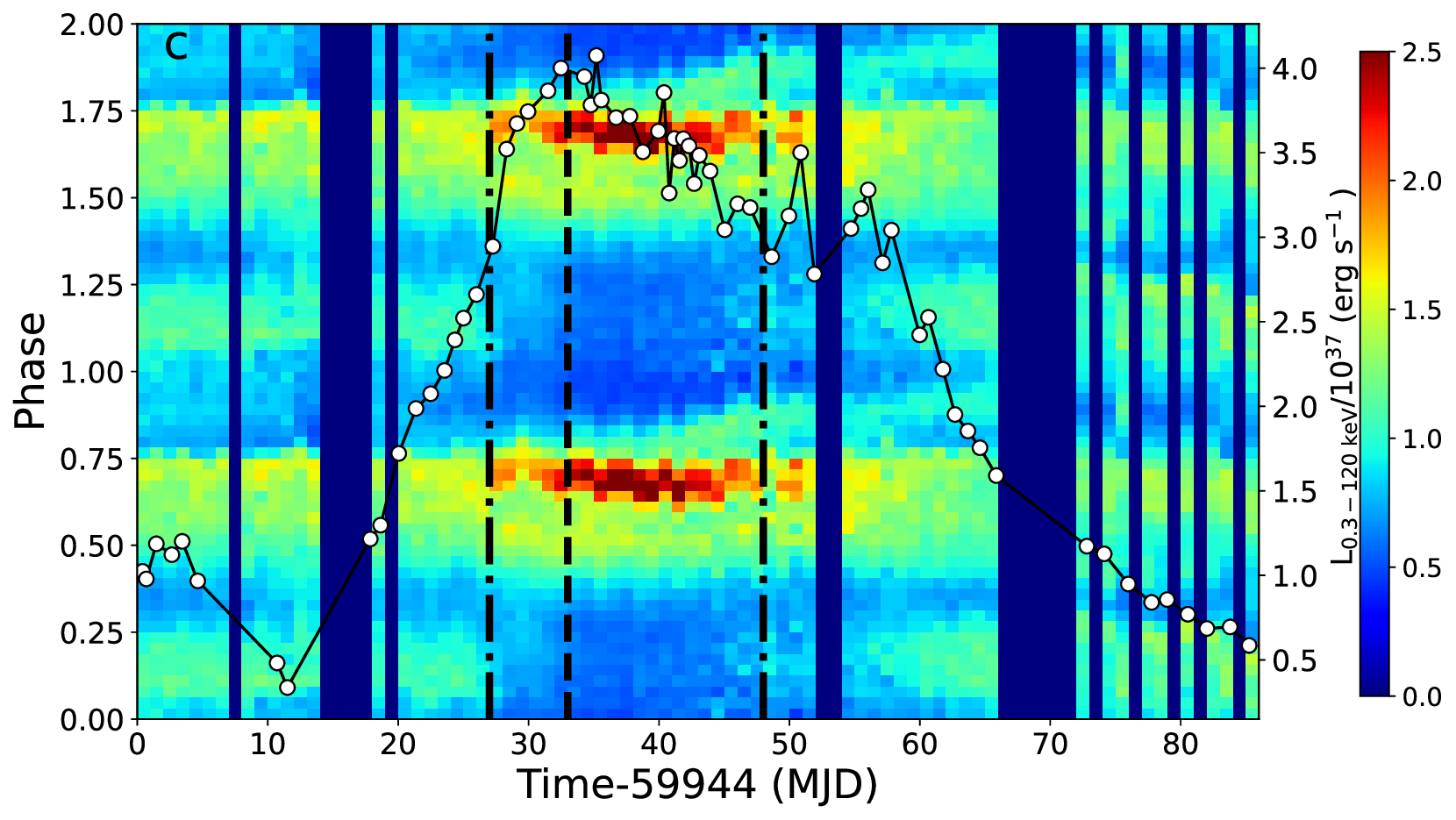}
    
    \caption{Panel a: Two-dimensional normalized pulse profiles of RX J0440.9+4431 in the 0.5--10\,keV energy band based on \textit{NICER} data. The color bar represents the intensity of the pulse, while the dark blue color indicates that it was not observed during this period. The white dots represent the X-ray luminosity in the energy range of 0.3--120\,keV, corresponding to the right y-axis. Two dashed dotted lines (MJD 59971; MJD 59992) indicate the transition of the pulse profile. The second dashed line (MJD 59977) marks the peak of the X-ray light curve for RX J0440.9+4431 in Fig.~\ref{fig:1}. Panel b: Two-dimensional normalized pulse profiles in the 10--30\,keV energy band based on the ME data. Panel c: Two-dimensional normalized pulse profiles in the 30--100\,keV energy band based using HE data.}
  \label{fig:3}
\end{figure}
%#####################################################################
\subsection{Evolution of the pulse profile}
\label{sec:3.2}
We analyze observations from both \textit{Insight}-HXMT and \textit{NICER} to study the evolution of the pulse profile of RX J0440.9+4431. Since there are no known orbital elements for the binary system, we do not apply any correction for orbital motion other than barycentric correction. Chi-squared statistics are used to evaluate the best frequency for the pulse of both \textit{Insight}-HXMT and \textit{NICER} observations \citep{1987Leahy}. Subsequently, we fold the net light curves for \textit{Insight}-HXMT observations and the events for \textit{NICER} observations to obtain the pulse profiles. Each pulse profile is divided into 32 bins and normalized by the average source intensity. The cross-correlation method is employed to align the pulse profiles with the reference pulse profile obtained from NICER's ObsID 5203610107 (with an energy range of 0.5--10\,keV and pepoch set to MJD 59942). 

It is widely acknowledged that the pulse profiles of X-ray pulsars are strongly dependent on both energy and luminosity. The resulting evolution of the observed pulse profiles with energy is presented in Fig.~\ref{fig:4}. The left column shows the observations at MJD 59947. From the low-energy pulse profile given by \textit{NICER}, we can see the dip structure reported in studies of the earlier outbursts at phase 0.81--0.87 \citep{2012Usui,2012Tsygankov}. \cite{2023Salganik} and \cite{2023Mandal} have also reported this result from their analysis of the \textit{NICER} data for this outburst. In the higher energy pulse profile given by \textit{Insight}-HXMT, this dip does not exist and the minor peak (phases 0.10--0.35) becomes more pronounced with increasing energy. The pulse profiles as a function of energy at the luminosity corresponding to MJD 59947 during the RX J0440.9+4431 decay (MJD 60016) are shown in the right column. Despite the complexity observed in the overall pulse profile compared to the left column, two peaks remain discernible. And the minor peak also demonstrates an increase in intensity with increasing energy. The middle column of Fig.~\ref{fig:4} shows the evolution of the pulse profiles when RX J0440.9+4431 is at its brightest. It is evident that the minor peak disappears, and the main peak (phases 0.45--0.75) splits into two wings with a phase separation of about 0.2. As the energy increases, the right wing becomes more prominent, which is consistent with the results presented by \cite{2023Salganik} using \textit{INTEGRAL} and \textit{NuSTAR} data. Furthermore, it is worth noting that in the middle column of Fig.~\ref{fig:4}, the high-energy (70--100\,keV) pulse profile peak and the low-energy (2--10\,keV) pulse profile peak are almost in phase. However, there is a phase difference between the high-energy peak and low-energy peak of the low-luminosity pulse profile (the left and right columns).

In Fig.~\ref{fig:3}, we present the phase-time matrix using the normalized pulse profiles for \textit{NICER}, \textit{Insight}-HXMT/ME and \textit{Insight}-HXMT/HE. However, the LE data from \textit{Insight}-HXMT were not included in the presentation due to their inadequate signal-to-noise ratio. The white dots in Fig.~\ref{fig:3} correspond to the luminosity in the energy range of 0.3--120\,keV on the right axis, which is calculated using {\tt cflux} and assuming a distance of 2.4\,kpc after fitting the model: {\tt Const}$\times${\tt TBabs}$\times${\tt (Gaussian}$+${\tt compTT}$+${\tt compTT)} (detailed results of \textit{Insight}-HXMT spectra analysis will be reported in the following paper, in preparation). The evolution of the pulse profile from the low-energy profile observed by \textit{NICER} to the high-energy profile obtained by \textit{Insight}-HXMT/HE is nearly consistent, as seen in Fig.~\ref{fig:3}. When the luminosity is below \textasciitilde$3\times10^{37}\ {\rm erg\ \rm s^{-1}}$, the pulse profile exhibits double peaks, while above \textasciitilde$3\times10^{37}\ {\rm erg\ \rm s^{-1}}$, it is single-peaked. Two dashed dotted lines (MJD 59971; MJD 59992) in Fig.~\ref{fig:3} correspond to the time points of the pulse profile transition. In panel c of Fig.~\ref{fig:1}, the HR also exhibits a transition at these same two dashed dotted lines. It is worth noting that when the luminosity is less than $3\times10^{37}\ {\rm erg\ \rm s^{-1}}$, the dip structure appears at a phase of about 0.8 in the soft energy pulse profile (Panel a of Fig.~\ref{fig:3}). Before MJD 59977, there was no dip in the hard energy range, consistent with previous reports. However, after MJD 59992, the main peak in the high-energy range also exhibited splitting at a phase of about 0.8 (Panel c of Fig.~\ref{fig:3}), which is difficult to explain using the explanations for the dip structure in the soft energy range.%, suggesting that it is more likely related to variations in the geometry of the accretion column.

\subsection{Evolution of the pulsed fraction}
\label{sec:3.3}
\begin{figure}
	\includegraphics[width=\columnwidth]{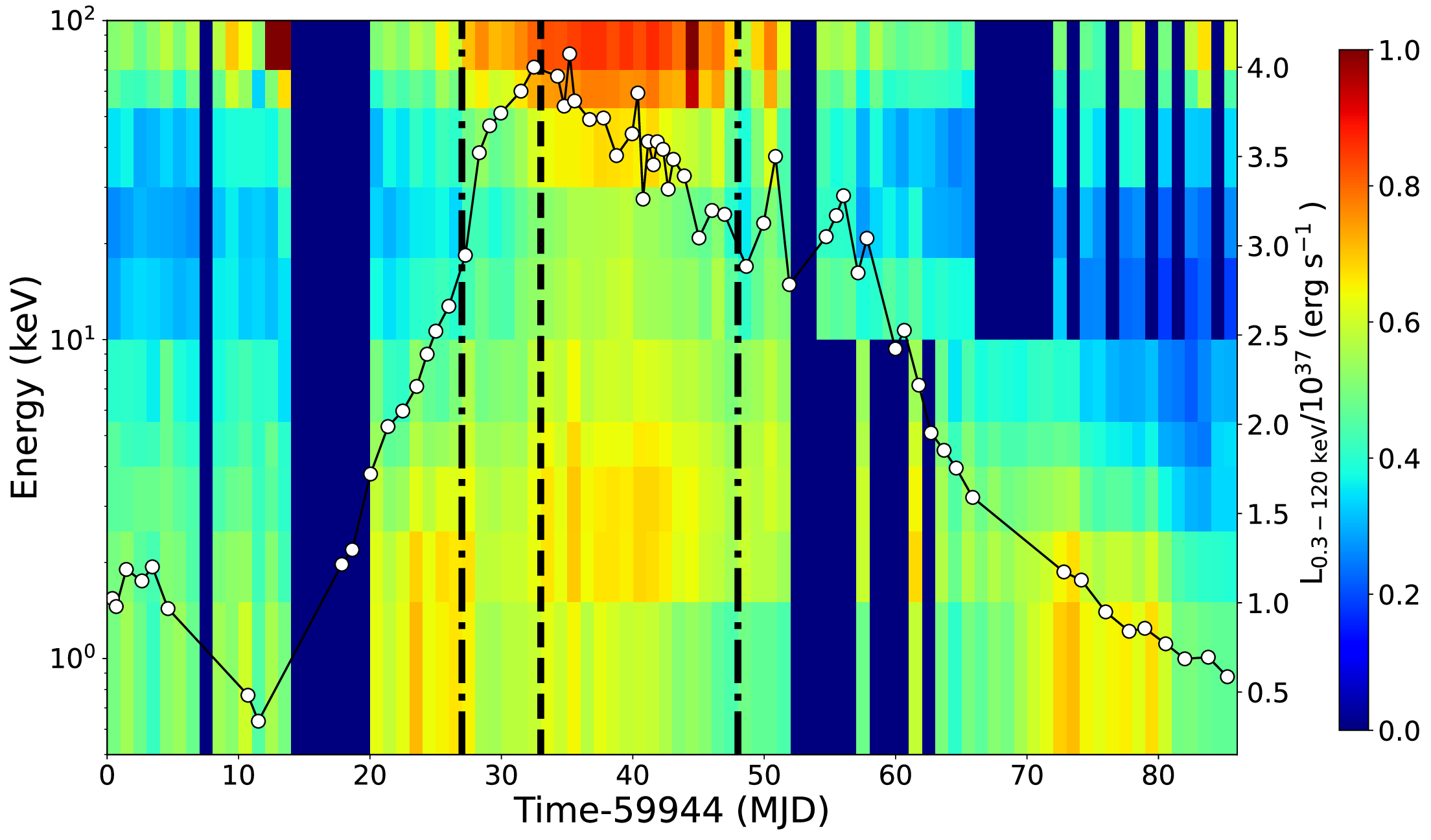}
    \caption{Two-dimensional pulsed fraction of RX J0440.9+4431 based on the \textit{NICER} and \textit{Insight}-HXMT. Data points below 10\,keV are from \textit{NICER}, while the other data points are from ME and HE. The energy range divisions are the same as shown in Fig.~\ref{fig:4}. The black dashed lines and white circles are the same as in Fig.~\ref{fig:3}. The points in extreme reddish-brown color due to using the net light curve to fold the pulse profile for \textit{Insight}-HXMT, resulting in negative counts and pulsar fraction larger than 1.}
  \label{fig:5}
\end{figure}

\begin{figure}
	\includegraphics[width=0.9\columnwidth]{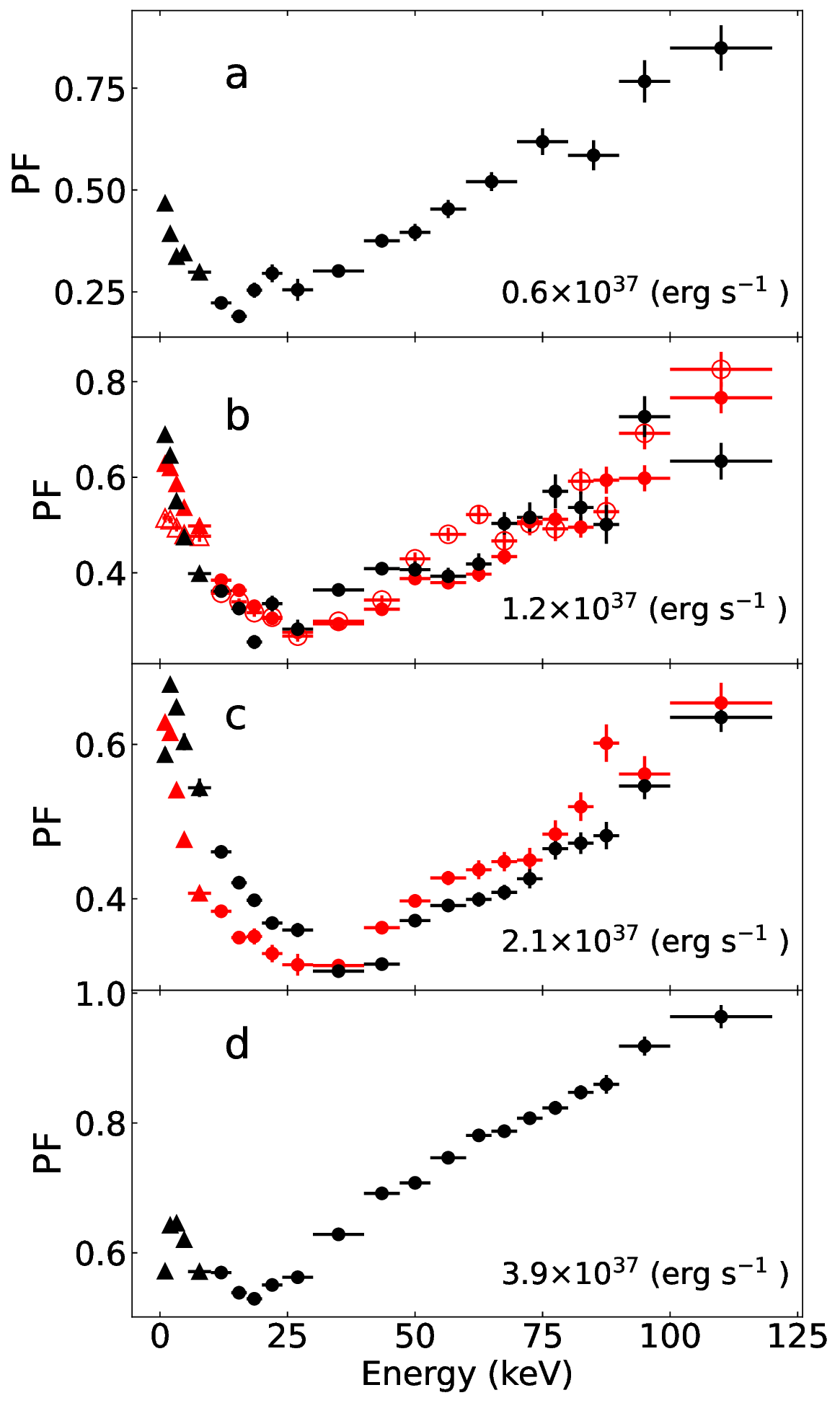}
    \caption{Pulsed fraction of RX J0440.9+4431 as a function of energy at different luminosities using \textit{NICER} (triangle mark) and \textit{Insight}-HXMT (circle mark) data. The red and black colors represent the rising and declining parts of the outburst, respectively. The hollow circles are derived from data prior to MID 59958.}
  \label{fig:6}
\end{figure}

\begin{figure*}
	\includegraphics[width=0.65\columnwidth]{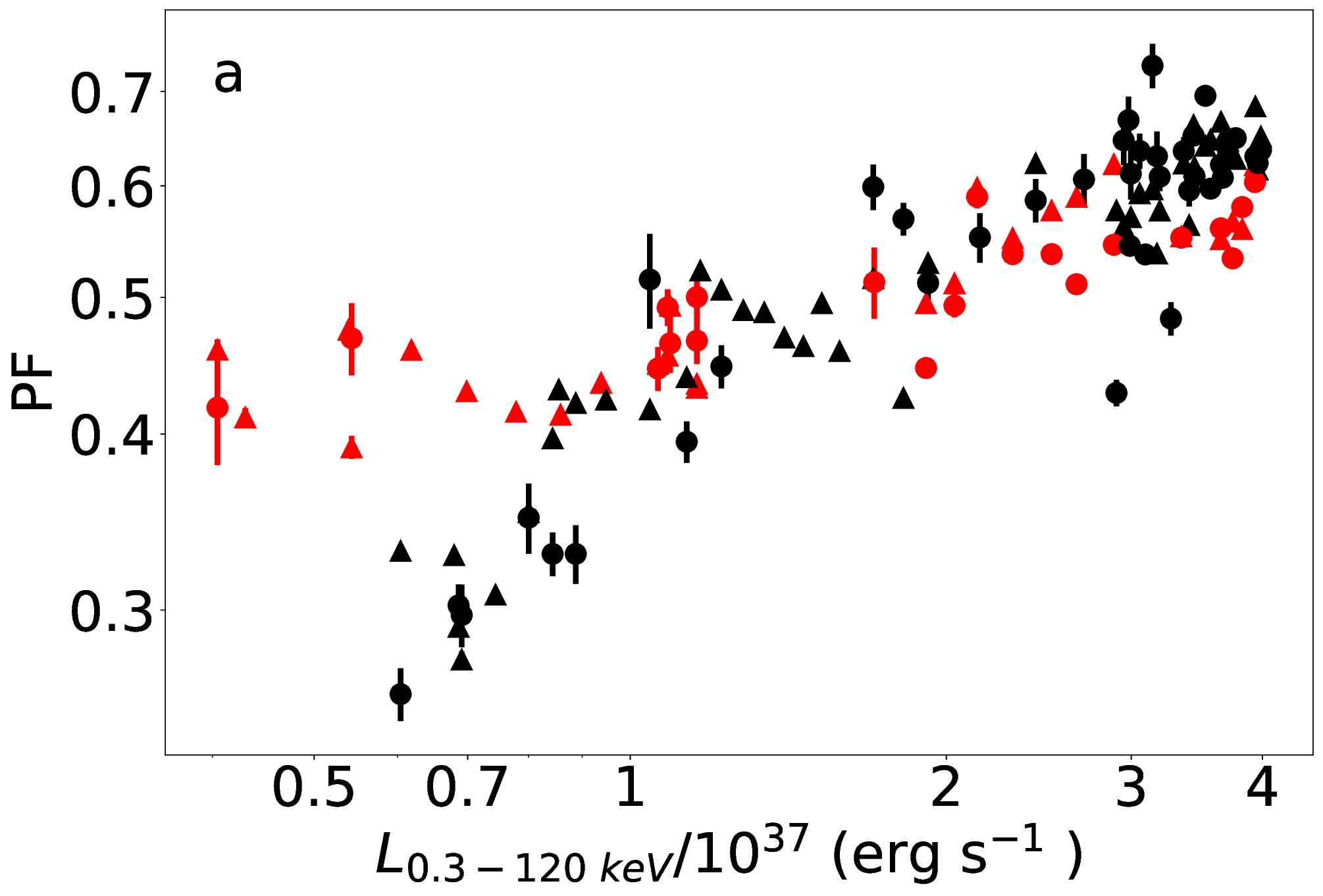}
    \includegraphics[width=0.65\columnwidth]{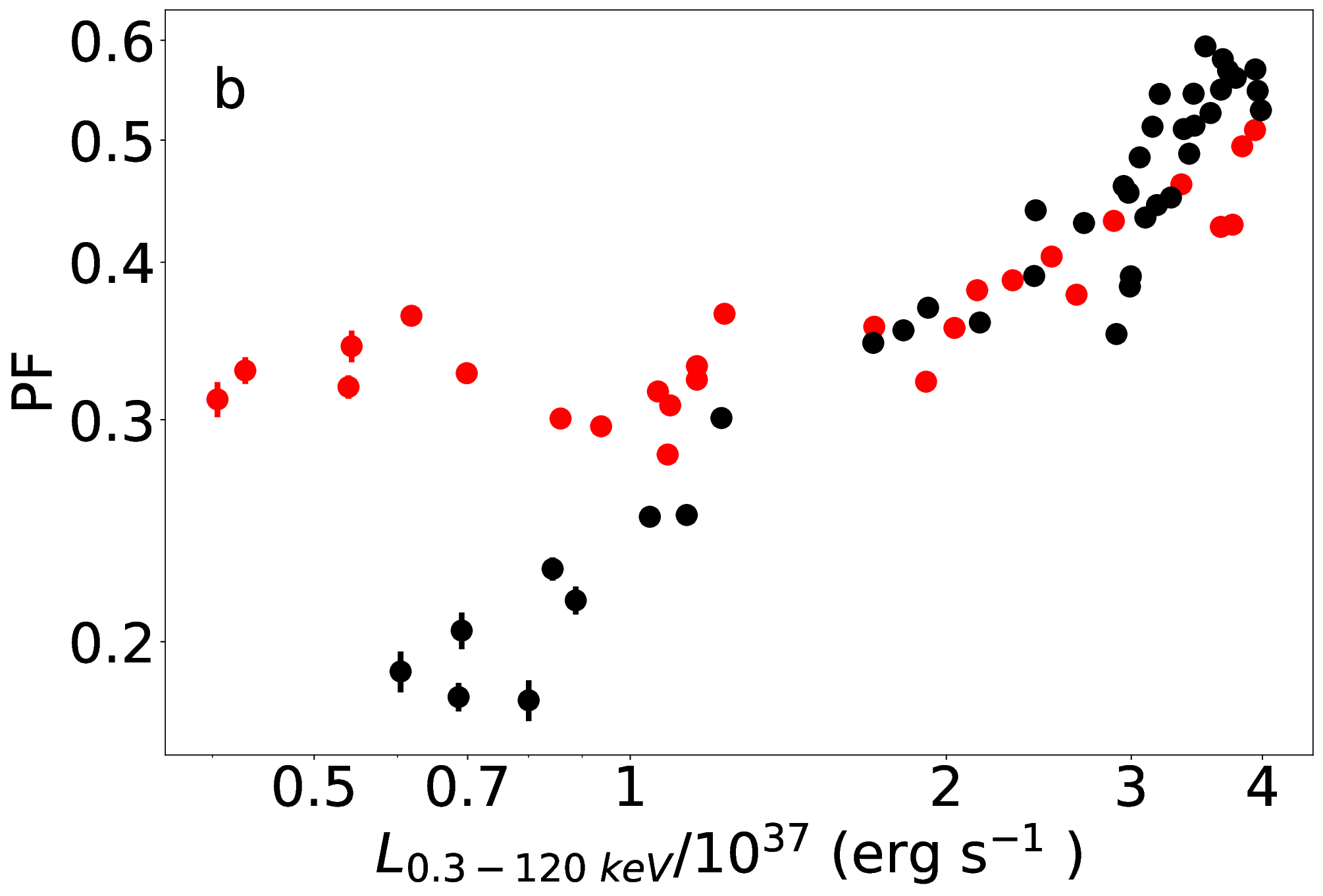}
    \includegraphics[width=0.65\columnwidth]{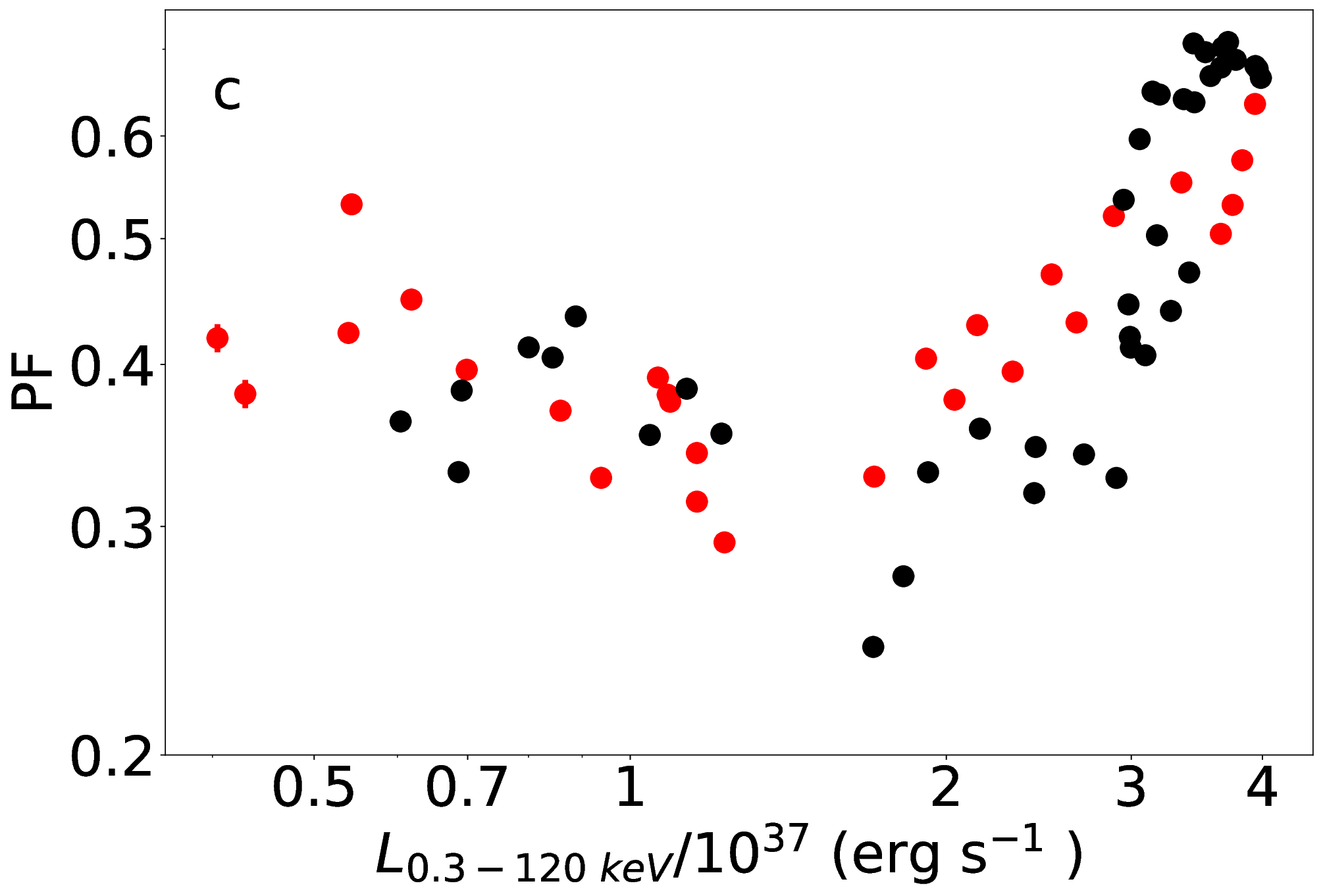}
    
    \caption{Panel a: Pulsed fraction of RX J0440.9+4431 as a function of luminosity in the 2--10\,keV energy range using \textit{NICER} (triangle mark) and \textit{Insight}-HXMT (circle mark) data. The red and black data points represent the data before and after the source reaches its peak luminosity, respectively. Panel b: Pulsed fraction in the 10--30\,keV energy range from ME. Panel c: Pulsed fraction in the 30--100\,keV energy range from HE.}
  \label{fig:7}
\end{figure*}

To quantify the pulse intensity, we calculated the pulsed fraction (PF) of RX J0440.9+4431:
\begin{equation}
\mathrm{PF}=\left(F_{\max }-F_{\min }\right) /\left(F_{\max }+F_{\min }\right),
\end{equation}
where $F_{\max }$ and $F_{\min }$ are the maximum and minimum fluxes in the pulse profile, respectively.
The energy-time matrix using the pulsed fraction for \textit{NICER} and \textit{Insight}-HXMT is shown in Fig.~\ref{fig:5}. The white dots and three black lines are the same as in Fig.~\ref{fig:1} and Fig.~\ref{fig:3}. During the outburst, the PF is at its lowest around \textasciitilde20--30\,keV, which aligns with the findings reported by \cite{2023Salganik} based on the analysis of \textit{NuSTAR} data. 
In Figure 5 of \cite{2023Salganik}, the PF for low luminosities does not appear as smooth around \textasciitilde60\,keV. Benefiting from the large effective area of \textit{Insight}-HXMT and its dense observations of RX J0440.9+4431, we have subdivided the energy ranges further, and Fig.~\ref{fig:6} depicts the variations in PF with energy for four different luminosity levels. In panels b and c, there is a less pronounced increase in the PF around \textasciitilde60\,keV. This behavior is not observed in panels a and d. Furthermore, the PF in panel d is larger than the other panels. This suggests a correlation between the PF and the luminosity \citep{2018Mushtukov,2020wang0243,2022Wang0535}. Therefore, we plot the variation of the PF with luminosity for RX J0440.9+4431 using \textit{NICER} and \textit{Insight}-HXMT data in Fig.~\ref{fig:7}. From panels a (2--10\,keV) and b (10--30\,keV) of Fig.~\ref{fig:7}, it can be seen that the PF is positively correlated with the luminosity. However, in the falling phase of the outburst, the relationship between the PF and luminosity is steeper compared to the rising phase. Moreover, for panel c (30--100\,keV) from the high-energy range, unlike panels a and b, the PF exhibits a slight negative correlation at low luminosity and a positive correlation at high luminosity, and the rising and falling phases of the outburst are similar.

%#####################################################################
%#####################################################################
%#####################################################################
%#####################################################################
\begin{figure}

    \includegraphics[width=\columnwidth]{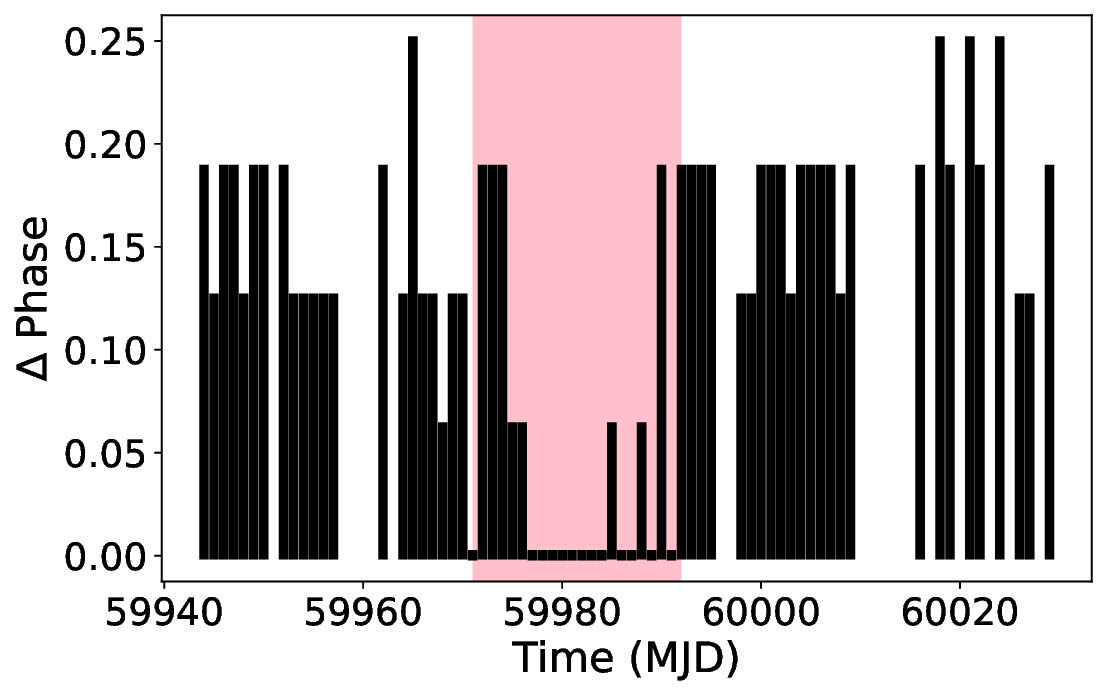}
    
    \caption{The temporal evolution of the phase difference between the high-energy (70--100\,keV) and low-energy (10--18\,keV) pulse profile peaks. The start and end positions of the red shaded region correspond to the dashed lines in Fig.~\ref{fig:3}.}
  \label{fig:8}
\end{figure}
%#####################################################################

\section{Discussion and Conclusion}
\label{sec:4}
We conduct an analysis of the \textit{NICER} and \textit{Insight}-HXMT data for the RX J0440.9+4431 giant outburst during 2022--2023, with a focus on the timing properties. The research further supplements the understanding of the timing properties in higher energy ranges. Prior to our study, \cite{2023Salganik} had analyzed the \textit{NICER} data and provided a comprehensive evolution of the low-energy pulse profile. They found that the pulse profile undergoes a transition from a double-peak pattern to a single-peak pattern at a luminosity of \textasciitilde$2.8\times10^{37}\ {\rm erg\ \rm s^{-1}}$, which was considered to be the critical luminosity. Additionally, \cite{2023Mandal} also found that the spectral parameters obtained from \textit{NICER} observations undergo changes at this luminosity. Above this luminosity, the correlation between the photon index and flux becomes positive, and the cut-off energy also increases with the increasing flux. We present a more complete evolution of the mid-energy (10--30\,keV) and high-energy (30--100\,keV) pulse profiles, revealing that they also transition from a double-peak pattern to a single-peak pattern around a luminosity of \textasciitilde$3\times10^{37}\ {\rm erg\ \rm s^{-1}}$, consistent with the low-energy evolution. The results suggest a significant alteration in the radiation beam, indicating the existence of different accretion regimes \citep{2013Reig}. These regimes are defined by a critical luminosity ($L_{\rm crit}$), which represents the luminosity at which the radiation pressure primarily begins to decelerates the accreting material above the neutron star. The changes in the radiation beam are typically accompanied by a change in the pulse profile shape and the spectral shape, as observed in other sources such as 1A 0535+262 \citep{2022Wang0535}, EXO 2030+375 \citep{2023Fu}, 4U 1901+03 \citep{2020Tuo}, RX J0209.6-7427 \citep{2022Hou}, and Swift J0243.6+6124 \citep{2022Liu}.

Additionally, benefiting from the high HXMT statistics and broad energy range, during high luminosity, we can observe that the peaks of the high-energy and low-energy pulse profiles coincide, indicating that they both originate from the same direction, as seen in the middle column of Fig.~\ref{fig:4}. Conversely, when the luminosity is low, the phase of the peak in the low-energy pulse profile differs from that of the peak in the high-energy pulse profile. This suggests that high-energy photons and low-energy photons are not emitted from the same direction. This feature was previously proposed in the pulse profile evolution of EXO 2030+375, where \cite{2023Fu} utilized the phase difference between the high-energy and low-energy pulse peaks to determine the critical luminosity during its 2021 giant outburst. To track the temporal evolution of the phase difference between the high-energy and low-energy pulse profile peaks in RX J0440.9+4431, we divide one pulse period into 16 bins to individually fold the profiles. This division helps mitigate occasional data point jumps caused by localized profile features. We compute $\Delta\ Phase$  corresponding to the maximum values of the high-energy (70--100\,keV) and low-energy (10--18\,keV) pulse profiles. The final results are presented in Fig.~\ref{fig:8}. From Fig.~\ref{fig:8}, we can see that within the red shaded region where the luminosity is greater than $3\times10^{37}\ {\rm erg\ \rm s^{-1}}$, $\Delta\ Phase$ is very small, and the peak of the high-energy profile aligns closely with the peak of the low-energy profile in phase. As discussed by \cite{2022Hou}, during high luminosity, both low-energy and high-energy photons mainly escape from the sides of the accretion column. There is a jump from 0 to 0.1875 at the start and end positions of the red shadow. Further examination of the pulse profile in the 10--18\,keV range reveal that the main peak splits into two wings, and these two wings have similar magnitudes. Therefore, during the transition from the left-dominant wing to the right-dominant wing in the pulse profile of the main peak, even slight instability can cause a phase jump. At higher luminosities, the low-energy pulse profile is completely dominated by the right-side swing, resulting in almost overlapping pulse peaks for high-energy and low-energy, as shown in Fig.~\ref{fig:4}. 
The critical luminosity obtained using this method is consistent with the results obtained from the transition of the pulse profile from single to double peaks and the evolution of the energy spectra. This further validates the feasibility of using this method to determine the critical luminosity.

\cite{2009Lutovinov} investigated the ten brightest X-ray pulsars using \textit{INTEGRAL} data in the hard ($> 20$\,keV) energy band and found that the pulse fraction increases with energy for all pulsars. Subsequently, similar findings of PF increasing with energy were reported in pulsars such as 2S 1845--024 \citep{2022Nabizadeh} \, GRO J2058+42 \citep{2022Gorban}, 1A 0535+262 \citep{2022Wang0535}, Swift J1808.4--1754 \citep{2022Salganik}, Swift J1626.6-5156 \citep{2021Molkov}, XTE J1858+034 \citep{2021Tsygankov}, 4U 1901+03 \citep{2021Raib}, eRASSU J050810.4--660653 \citep{2022Salganika}, Swift J1816.7--1613 \citep{2019Nabizadeh}, etc. High-energy photons have a significantly larger optical depth along the magnetic field direction, allowing them to escape only from the sides of the accretion column. In contrast, low-energy photons have less restricted escape directions and can exit both from the sides and the top of the column \citep{2022Hou}. This leads to a more concentrated emission for higher-energy radiation compared to lower-energy radiation, which provides a possible explanation for the increase in pulse fraction with energy for high-energy photons. In addition, \cite{2009Lutovinov} also found that in many cases, this increase is not monotonic, exhibiting local features near the cyclotron line harmonics. As shown in Fig.~\ref{fig:5} and Fig.~\ref{fig:6}, there is a prominent concave in the PF at \textasciitilde20--30\,keV for RX J0440.9+4431, but no cyclotron absorption lines were detected in its spectra \citep{2023Salganik}. During the outburst in September 2010, \cite{2012Tsygankov} reported the discovery of a possible cyclotron absorption line at 32\,keV in RX J0440.9+4431. Notably, Fig.~\ref{fig:6} reveals the PF does not appear to increase smoothly with energy at the first harmonic of the cyclotron absorption line. However, when we use the RMS pulse fraction calculated using equation A.3 from \cite{2023Ferrigno} to quantify the pulse profile, this feature becomes less prominent (Fig.~\ref{fig:A}). Consequently, it is difficult to consider the possibility that these features are attributed to the effects of resonant absorption. It may be worth considering the use of the more precise PF spectrum proposed by \cite{2023Ferrigno} in the subsequent work. At the low luminosity, 1A 0535+262 exhibits a prominent concave in the PF around 20--30\,keV, which \cite{2023Chhotaray} attributed this concave to the diffused X-ray emission from the relatively hot plasma above the pole of the neutron star. \cite{2015Postnov} and \cite{2021Kylafis} studied the reflection of continuum X-ray photons at the atmosphere of a magnetic neutron star. At low and high energies, the reflected spectra are significantly reduced due to absorption and downward scattering, resulting in a prominent bump around 20--30\,keV. The process of reflection redistributes the emitted photons, leading us to hypothesize that the decreased PF at \textasciitilde20--30\,keV may arise from the dilution of reflected photons. The proportion of reflected photons affects the depth of the concave, and this proportion is related to the height of the accretion column \citep{2013Poutanen}. A higher accretion column results in a lower probability of photon scattering. Above the critical luminosity, the concave is weaker, which may suggest a lower proportion of reflected photons and a higher accretion column compared to the low luminosity.

As shown above, the pulse profiles and pulse fractions of X-ray pulsars exhibit a dependency on their luminosity. In Fig.~\ref{fig:7}, a transition in the pulse fraction variation with luminosity occurs at a luminosity of \textasciitilde$2\times10^{37}\ {\rm erg\ \rm s^{-1}}$. This luminosity is very close to the critical luminosity defined in our study. Above this luminosity, the pulse fraction increases with increasing luminosity for high-energy, medium-energy, and low-energy photons. Below this luminosity, the pulse fraction for high-energy photons shows consistent variations with luminosity in the rising and falling phases. However, for low-energy photons, the pulse fraction in the falling phase is lower than that in the rising phase. This discrepancy can be attributed to the fact that while high-energy photons predominantly originate from the polar caps via direct emission, low-energy photons comprise contributions from direct emission, thermal components, and downward scattering of high-energy photons. This implies that the generation process of low-energy photons may differ between the rising and falling phases. Inconsistencies in the physical parameters at the same luminosity during the rising and falling phases have been observed in other sources, such as 1A 0535+262 \citep{2021Kong} and V 0332+53 \citep{2017Doroshenko}, in terms of cyclotron absorption line energies. These discrepancies have been explained as possibly being related to the geometric shape of the radiation region during the rising and falling phases, potentially caused by changes in the accretion disk structure and its interaction with the magnetosphere of the neutron star. However, the exact cause of these differences is not yet well understood.

Currently, our understanding of the accretion physics in X-ray pulsars is still limited. The pulse profile is influenced by various factors such as the viewing angle to the magnetic pole, the distribution of radiation beams, and relativistic effects. The temporal properties of X-ray accreting pulsars require further investigation through detailed simulations and higher-quality observational data in the future.

\section*{Acknowledgements}

This research utilized data and software from the High Energy Astrophysics Science Archive Research Center (HEASARC), provided by NASA's Goddard Space Flight Center, as well as the \textit{Insight}-HXMT mission, supported by the China National Space Administration (CNSA) and the Chinese Academy of Sciences (CAS). This work is supported by the National Key R\&D Program of China (2021YFA0718500). We acknowledge funding support from the National Natural Science Foundation of China (NSFC) under grant No. 12122306, No. U2038102, No. U2031205, No. U2038104, No. U1838201, No. U1838108, No. 12173103, No. 12041303, the CAS Pioneer Hundred Talent Program Y8291130K2 and the Scientific and technological innovation project of IHEP Y7515570U1.

%%%%%%%%%%%%%%%%%%%%%%%%%%%%%%%%%%%%%%%%%%%%%%%%%%
\section*{Data Availability}

 The data of \textit{insight}-HXMT can be obtained from this website \url{http://hxmtweb.ihep.ac.cn/}. The data of \textit{NICER} used for this paper are publicly available in the High Energy Astrophysics Science Archive Research Centre (HEASARC) at \url{https://heasarc.gsfc.nasa.gov/cgi-bin/W3Browse/w3browse.pl}.



\bibliographystyle{mnras}
\bibliography{example} % if your bibtex file is called example.bib

%\begin{thebibliography}{99}
%\end{thebibliography}

%%%%%%%%%%%%%%%%%%%%%%%%%%%%%%%%%%%%%%%%%%%%%%%%%%

%%%%%%%%%%%%%%%%% APPENDICES %%%%%%%%%%%%%%%%%%%%%

\appendix
\section{Figures}
\label{A}
\begin{figure}
	\includegraphics[width=0.9\columnwidth]{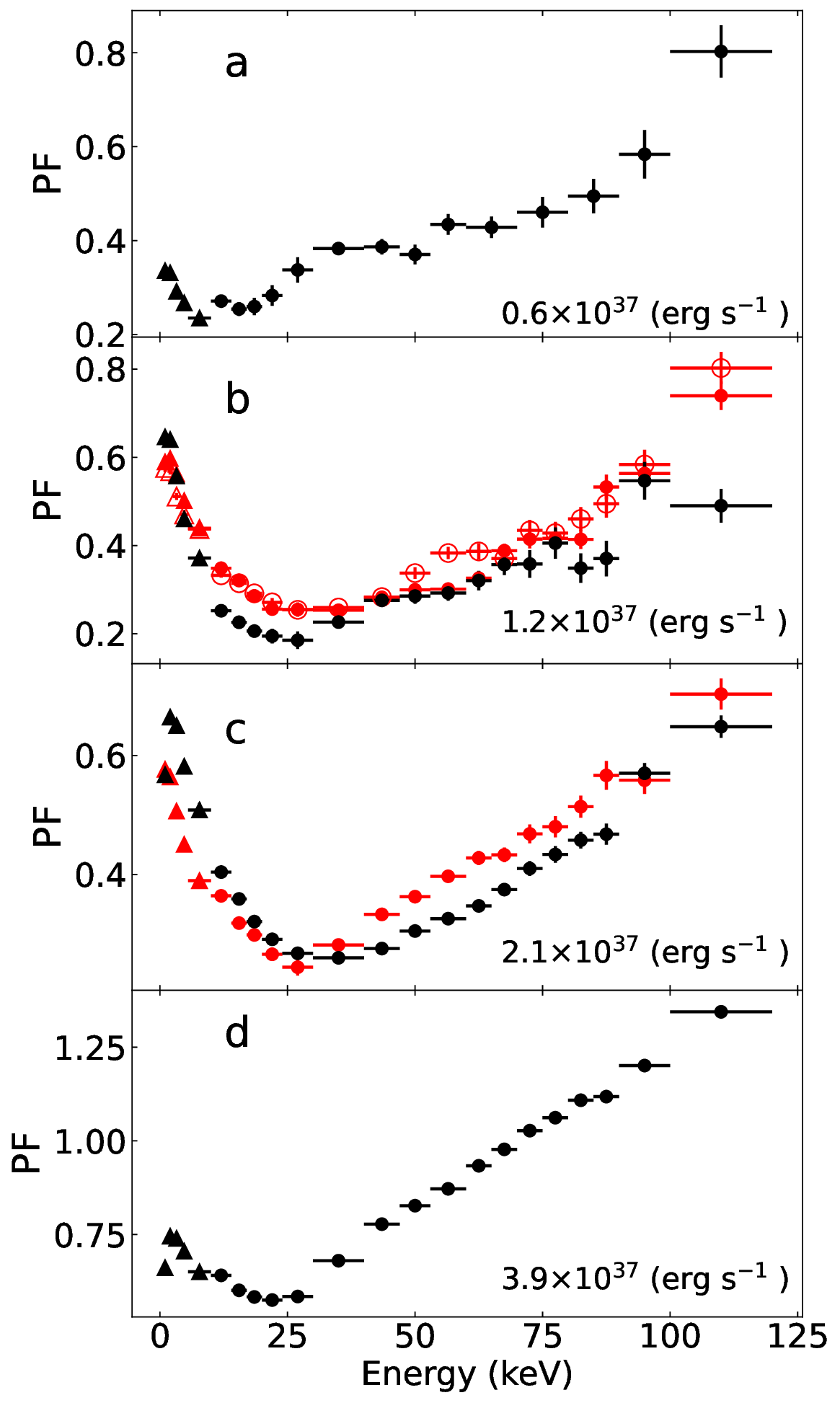}
    \caption{The RMS pulse fraction of RX J0440.9+4431 as a function of energy at different luminosities using \textit{NICER} (triangle mark) and \textit{Insight}-HXMT (circle mark) data. The red and black colors represent the rising and declining parts of the outburst, respectively. The hollow circles are derived from data prior to MID 59958.}
  \label{fig:A}
\end{figure}

%%%%%%%%%%%%%%%%%%%%%%%%%%%%%%%%%%%%%%%%%%%%%%%%%
\section{Tables}
\label{B}
\begin{center}
\onecolumn
\setlength{\tabcolsep}{2mm}
\begin{longtable}{cccccc}

    \caption{\textit{Insight}-HXMT observations log used in this paper. The net count rates in the HE band (30$-$100\,keV) and the periods are listed. }
    
    \label{tab:hxmt}\\
    
    \hline
    ObsIDs&ExpIDs & Start Time (MJD) & HE Exposure Time (s)& HE Count Rate (cts $\rm{s^{-1}}$) &Period (s) \\
    \hline   
    \endfirsthead
    \hline
    ObsIDs&ExpIDs & Start time (MJD) & Exposure time (s)&Count Rate (cts $\rm{s^{-1}}$)  &Period (s) \\
    \hline   
    \endhead
    \hline 
    \multicolumn{5}{r}{to be continued}\\
\hline
    \endfoot
    \hline
    \endlastfoot
    
P0514361001 & P051436100101 & 59944.40 & 1971 & $181.1\pm{0.3}$ & $208.73\pm{1.80}$\\
 & P051436100102 & 59944.56 & 4298 & $177.5\pm{0.2}$ \\
 & P051436100104 & 59944.82 & 3290 & $204.9\pm{0.2}$ & $208.01\pm{0.67}$\\
 & P051436100105 & 59944.96 & 2389 & $175.2\pm{0.3}$ \\
 \hline
P0514361002 & P051436100201 & 59945.46 & 4467 & $210.3\pm{0.2}$ & $207.98\pm{0.55}$\\
 & P051436100202 & 59945.62 & 2082 & $200.3\pm{0.3}$ \\
 & P051436100203 & 59945.75 & 578 & $180.2\pm{0.6}$ \\
 \hline
P0514361003 & P051436100301 & 59946.65 & 1125 & $203.7\pm{0.4}$ & $207.87\pm{0.48}$\\
 & P051436100302 & 59946.81 & 3444 & $208.6\pm{0.2}$ \\
 & P051436100303 & 59946.94 & 2351 & $180.6\pm{0.3}$ \\
 \hline
P0514361004 & P051436100401 & 59947.44 & 5129 & $206.6\pm{0.2}$ & $208.03\pm{0.45}$\\
 & P051436100402 & 59947.60 & 169 & $258.4\pm{1.2}$ \\
 & P051436100403 & 59947.73 & 741 & $213.9\pm{0.5}$ \\
 \hline
P0514361005 & P051436100501 & 59948.63 & 1636 & $172.7\pm{0.3}$ & $207.84\pm{0.49}$\\
 & P051436100502 & 59948.79 & 3381 & $181.2\pm{0.2}$ \\
 & P051436100503 & 59948.92 & 2296 & $189.7\pm{0.3}$ \\
 \hline
P0514361008 & P051436100801 & 59949.95 & 4681 & $160.1\pm{0.2}$ & $207.79\pm{0.63}$\\
 & P051436100802 & 59950.11 & 1789 & $164.9\pm{0.3}$ \\
 & P051436100803 & 59950.24 & 1001 & $110.8\pm{0.3}$ \\
 & P051436100804 & 59950.37 & 3932 & $129.3\pm{0.2}$ & $207.92\pm{0.45}$\\
 & P051436100805 & 59950.51 & 2623 & $141.7\pm{0.2}$ \\
 & P051436100806 & 59950.64 & 1189 & $133.1\pm{0.3}$ \\
 & P051436100807 & 59950.77 & 3276 & $164.8\pm{0.2}$ & $207.84\pm{0.47}$\\
 & P051436100808 & 59950.91 & 3420 & $138.5\pm{0.2}$ \\
 & P051436100809 & 59951.04 & 2976 & $154.4\pm{0.2}$ \\
 \hline
P0514361009 & P051436100901 & 59952.20 & 698 & $124.3\pm{0.4}$ & $208.00\pm{0.64}$\\
 & P051436100902 & 59952.36 & 4083 & $122.7\pm{0.2}$ \\
 & P051436100903 & 59952.49 & 1511 & $93.7\pm{0.2}$ & $208.04\pm{2.13}$\\
 & P051436100904 & 59952.63 & 522 & $115.5\pm{0.5}$ \\
 \hline
P0514361015 & P051436101501 & 59953.13 & 1227 & $86.9\pm{0.3}$ & $207.91\pm{1.65}$\\
 & P051436101502 & 59953.28 & 2026 & $98.3\pm{0.2}$ \\
 & P051436101503 & 59953.42 & 2768 & $90.5\pm{0.2}$ \\
 \hline
P0514361016 & P051436101601 & 59954.72 & 2983 & $79.7\pm{0.2}$ & $207.39\pm{2.01}$\\
 & P051436101602 & 59954.87 & 3041 & $59.2\pm{0.1}$ \\
 & P051436101603 & 59955.01 & 1554 & $74.3\pm{0.2}$ \\
 \hline
P0514361017 & P051436101701 & 59955.51 & 576 & $51.0\pm{0.3}$ & $207.91\pm{2.02}$\\
 & P051436101702 & 59955.67 & 2578 & $65.0\pm{0.2}$ \\
 & P051436101703 & 59955.80 & 1526 & $50.3\pm{0.2}$ \\
 \hline
P0514361018 & P051436101801 & 59956.70 & 3657 & $38.9\pm{0.1}$ & $207.59\pm{1.08}$\\
 & P051436101802 & 59956.86 & 1473 & $44.2\pm{0.2}$ \\
 & P051436101803 & 59956.98 & 929 & $51.6\pm{0.2}$ \\
 \hline
P0514361019 & P051436101901 & 59957.63 & 3311 & $33.1\pm{0.1}$ & ~.~.~.\\
 & P051436101902 & 59957.78 & 2841 & $35.7\pm{0.1}$ \\
 & P051436101903 & 59957.91 & 1398 & $31.1\pm{0.1}$ \\
 \hline
P0514361022 & P051436102201 & 59961.86 & 1403 & $218.1\pm{0.4}$ &~.~.~.\\
 & P051436102202 & 59962.00 & 584 & $201.8\pm{0.6}$ \\
 & P051436102203 & 59962.13 & 1344 & $255.0\pm{0.4}$ \\
 \hline
P0514361023 & P051436102301 & 59962.65 & 3289 & $262.2\pm{0.3}$ & $207.70\pm{0.44}$\\
 & P051436102302 & 59962.81 & 1941 & $255.4\pm{0.4}$ \\
 & P051436102303 & 59962.94 & 691 & $284.6\pm{0.6}$ \\
 \hline
P0514361024 & P051436102401 & 59964.04 & 925 & $397.7\pm{0.7}$ & $207.65\pm{0.58}$\\
 & P051436102402 & 59964.19 & 3989 & $368.9\pm{0.3}$ \\
 & P051436102403 & 59964.33 & 947 & $421.9\pm{0.7}$ \\
 \hline
P0514361025 & P051436102501 & 59965.37 & 871 & $454.3\pm{0.7}$ & $207.59\pm{0.48}$\\
 & P051436102502 & 59965.52 & 2015 & $426.3\pm{0.5}$ \\
 & P051436102503 & 59965.65 & 2240 & $429.8\pm{0.4}$ \\
 \hline
P0514361026 & P051436102601 & 59966.50 & 2315 & $519.6\pm{0.5}$ & $207.58\pm{0.55}$\\
 & P051436102602 & 59966.64 & 2193 & $442.7\pm{0.4}$ \\
 & P051436102603 & 59966.77 & 803 & $441.2\pm{0.7}$ \\
 \hline
P0514361027 & P051436102701 & 59967.56 & 2248 & $506.6\pm{0.5}$ & $207.59\pm{0.58}$\\
 & P051436102702 & 59967.70 & 1619 & $535.6\pm{0.6}$ \\
 & P051436102703 & 59967.83 & 399 & $531.1\pm{1.2}$ \\
 \hline
P0514361028 & P051436102801 & 59968.35 & 1068 & $540.1\pm{0.7}$ & $207.52\pm{0.42}$\\
 & P051436102802 & 59968.49 & 2227 & $549.0\pm{0.5}$ \\
 & P051436102803 & 59968.62 & 2134 & $567.6\pm{0.5}$ \\
 \hline
P0514361029 & P051436102901 & 59969.01 & 1757 & $569.1\pm{0.6}$ & $207.43\pm{0.21}$\\
 & P051436102902 & 59969.14 & 3288 & $594.2\pm{0.4}$ \\
 & P051436102904 & 59969.42 & 2185 & $565.1\pm{0.5}$ \\
 & P051436102905 & 59969.55 & 2482 & $603.5\pm{0.5}$ \\
 & P051436102906 & 59969.68 & 2061 & $593.5\pm{0.5}$ \\
 \hline
P0514361030 & P051436103001 & 59970.00 & 2144 & $649.2\pm{0.6}$ & $207.39\pm{0.26}$\\
 & P051436103002 & 59970.14 & 3197 & $618.7\pm{0.4}$ \\
 & P051436103005 & 59970.54 & 2518 & $638.9\pm{0.5}$ \\
 & P051436103006 & 59970.68 & 2009 & $745.7\pm{0.6}$ \\
 \hline
P0514361031 & P051436103101 & 59971.26 & 571 & $719.2\pm{1.1}$ & $207.36\pm{0.41}$\\
 & P051436103102 & 59971.41 & 2415 & $728.4\pm{0.5}$ \\
 & P051436103103 & 59971.54 & 2657 & $776.3\pm{0.5}$ \\
 & P051436103104 & 59971.67 & 1163 & $732.3\pm{0.8}$ \\
 \hline
P0514361032 & P051436103201 & 59972.32 & 7350 & $812.7\pm{0.3}$ & $207.31\pm{0.43}$\\
 & P051436103202 & 59972.70 & 601 & $843.1\pm{1.2}$ \\
 \hline
P0514361035 & P051436103501 & 59973.11 & 2296 & $844.8\pm{0.6}$ & $207.21\pm{0.20}$\\
 & P051436103502 & 59973.25 & 7933 & $846.7\pm{0.3}$ \\
 & P051436103503 & 59973.69 & 918 & $909.6\pm{1.0}$ \\
 & P051436103504 & 59973.82 & 542 & $882.9\pm{1.3}$ \\
 & P051436103505 & 59973.96 & 4329 & $844.7\pm{0.4}$ & $207.12\pm{0.19}$\\
 & P051436103506 & 59974.10 & 2277 & $832.5\pm{0.6}$ \\
 & P051436103507 & 59974.24 & 13379 & $848.9\pm{0.3}$ \\
 & P051436103508 & 59974.69 & 3740 & $840.5\pm{0.5}$ \\
 \hline
P0514361036 & P051436103601 & 59975.49 & 6608 & $849.9\pm{0.4}$ & $206.99\pm{0.28}$\\
 & P051436103602 & 59975.68 & 3068 & $851.9\pm{0.5}$ \\
 & P051436103603 & 59975.80 & 446 & $965.9\pm{1.5}$ \\
 & P051436103604 & 59975.94 & 5930 & $824.9\pm{0.4}$ \\
 \hline
P0514361037 & P051436103701 & 59976.48 & 6960 & $829.6\pm{0.3}$ & $206.85\pm{0.23}$\\
 & P051436103702 & 59976.67 & 3084 & $849.1\pm{0.5}$ \\
 & P051436103703 & 59976.79 & 538 & $820.1\pm{1.2}$ \\
 & P051436103704 & 59976.93 & 6110 & $848.8\pm{0.4}$ \\
 & P051436103705 & 59977.08 & 2106 & $878.5\pm{0.6}$ \\
 \hline
P0514361038 & P051436103801 & 59977.80 & 3828 & $876.5\pm{0.5}$ & $206.78\pm{0.41}$\\
 & P051436103802 & 59978.00 & 4776 & $839.9\pm{0.4}$ \\
 & P051436103803 & 59978.14 & 549 & $645.6\pm{1.1}$ \\
 & P051436103804 & 59978.27 & 13297 & $837.6\pm{0.3}$ & $206.70\pm{0.32}$\\
 & P051436103805 & 59978.65 & 3113 & $828.4\pm{0.5}$ \\
 & P051436103806 & 59978.78 & 833 & $781.5\pm{1.0}$ & $206.66\pm{0.48}$\\
 & P051436103807 & 59978.92 & 6286 & $860.1\pm{0.4}$ \\
 & P051436103808 & 59979.06 & 1285 & $813.1\pm{0.8}$ \\
 & P051436103809 & 59979.20 & 10514 & $838.8\pm{0.3}$ & $206.61\pm{0.45}$\\
 & P051436103810 & 59979.57 & 3059 & $781.1\pm{0.5}$ & $206.58\pm{0.27}$\\
 & P051436103811 & 59979.70 & 1482 & $886.5\pm{0.8}$ \\
 & P051436103812 & 59979.83 & 4536 & $811.8\pm{0.4}$ \\
 & P051436103813 & 59979.98 & 3618 & $846.4\pm{0.5}$ \\
 & P051436103814 & 59980.12 & 950 & $782.7\pm{0.9}$ \\
 \hline
P0514361040 & P051436104001 & 59980.71 & 2681 & $824.9\pm{0.6}$ & $206.46\pm{0.29}$\\
 & P051436104002 & 59980.90 & 5392 & $832.0\pm{0.4}$ \\
 & P051436104003 & 59981.04 & 19 & $654.3\pm{5.9}$ \\
 & P051436104004 & 59981.18 & 2738 & $813.6\pm{0.5}$ \\
 \hline
P0514361041 & P051436104101 & 59981.77 & 3100 & $804.5\pm{0.5}$ & $206.40\pm{0.29}$\\
 & P051436104102 & 59981.96 & 1521 & $854.8\pm{0.7}$ \\
 & P051436104103 & 59982.10 & 3514 & $746.7\pm{0.5}$ \\
 \hline
P0514361042 & P051436104201 & 59982.76 & 2674 & $861.8\pm{0.6}$ & $206.34\pm{0.32}$\\
 & P051436104202 & 59982.95 & 1082 & $680.0\pm{0.8}$ \\
 & P051436104203 & 59983.09 & 3452 & $719.2\pm{0.5}$ \\
 \hline
P0514361043 & P051436104301 & 59983.95 & 670 & $734.2\pm{1.0}$ & $206.28\pm{0.39}$\\
 & P051436104302 & 59984.08 & 2011 & $786.1\pm{0.6}$ \\
 & P051436104303 & 59984.25 & 2456 & $758.5\pm{0.6}$ \\
 & P051436104304 & 59984.39 & 2557 & $745.1\pm{0.5}$ & $206.18\pm{0.44}$\\
 & P051436104305 & 59984.52 & 2864 & $763.7\pm{0.5}$ \\
 & P051436104307 & 59984.78 & 2609 & $725.3\pm{0.5}$ & $206.29\pm{0.52}$\\
 & P051436104308 & 59984.91 & 415 & $637.2\pm{1.2}$ \\
 & P051436104309 & 59985.05 & 734 & $730.9\pm{1.0}$ \\
 & P051436104310 & 59985.18 & 2484 & $734.5\pm{0.5}$ & $206.21\pm{0.41}$\\
 & P051436104311 & 59985.31 & 2484 & $771.7\pm{0.6}$ \\
 & P051436104312 & 59985.44 & 2711 & $707.6\pm{0.5}$ \\
 & P051436104313 & 59985.57 & 1497 & $776.1\pm{0.7}$ & $206.16\pm{0.67}$\\
 & P051436104314 & 59985.71 & 1445 & $754.2\pm{0.7}$ \\
 \hline
P0514361044 & P051436104401 & 59985.87 & 1481 & $766.9\pm{0.7}$ & $206.15\pm{0.42}$\\
 & P051436104402 & 59986.00 & 193 & $728.0\pm{1.9}$ \\
 & P051436104403 & 59986.13 & 2895 & $812.9\pm{0.5}$ \\
 & P051436104404 & 59986.30 & 2479 & $747.6\pm{0.5}$ & $206.11\pm{0.48}$\\
 & P051436104405 & 59986.43 & 2731 & $751.7\pm{0.5}$ \\
 & P051436104406 & 59986.56 & 1442 & $700.8\pm{0.7}$ \\
 & P051436104407 & 59986.70 & 1449 & $743.0\pm{0.7}$ & $206.11\pm{0.33}$\\
 & P051436104408 & 59986.83 & 1124 & $760.1\pm{0.8}$ \\
 & P051436104410 & 59987.09 & 240 & $691.4\pm{1.7}$ \\
 & P051436104411 & 59987.22 & 1371 & $730.1\pm{0.7}$ \\
 \hline
P0514361045 & P051436104502 & 59988.05 & 2041 & $702.3\pm{0.6}$ & $206.10\pm{0.45}$\\
 & P051436104503 & 59988.21 & 1518 & $780.7\pm{0.7}$ \\
 \hline
P0514361046 & P051436104602 & 59989.04 & 1357 & $729.1\pm{0.7}$ & $205.84\pm{1.17}$\\
 \hline
P0514361047 & P051436104702 & 59990.03 & 3149 & $692.7\pm{0.5}$ & $206.00\pm{1.01}$\\
 \hline
P0514361048 & P051436104802 & 59990.99 & 1923 & $742.7\pm{0.6}$ & $205.88\pm{1.84}$\\
 & P051436104803 & 59991.12 & 2569 & $664.2\pm{0.5}$ & $205.88\pm{0.89}$\\
 & P051436104804 & 59991.25 & 1555 & $660.4\pm{0.7}$ \\
 \hline
P0514361049 & P051436104902 & 59992.64 & 2913 & $633.8\pm{0.5}$ & $205.54\pm{0.92}$\\
 & P051436104903 & 59992.77 & 565 & $765.8\pm{1.2}$ \\
 \hline
P0514361050 & P051436105002 & 59993.96 & 2217 & $746.8\pm{0.6}$ & $205.65\pm{0.61}$\\
 & P051436105003 & 59994.09 & 2607 & $627.9\pm{0.5}$ \\
 \hline
P0514361051 & P051436105101 & 59994.86 & 1006 & $683.8\pm{0.8}$ & $205.77\pm{1.61}$\\
 \hline
P0514361052 & P051436105201 & 59995.92 & 2775 & $607.4\pm{0.5}$ & $205.63\pm{0.54}$\\
 & P051436105202 & 59996.07 & 2661 & $645.2\pm{0.5}$ \\
 \hline
P0514361053 & P051436105302 & 59998.19 & 2225 & $652.6\pm{0.5}$ & $205.43\pm{0.26}$\\
 & P051436105305 & 59998.58 & 2640 & $659.9\pm{0.5}$ \\
 & P051436105307 & 59998.85 & 2454 & $598.6\pm{0.5}$ & $205.45\pm{0.38}$\\
 & P051436105308 & 59998.98 & 2948 & $658.8\pm{0.5}$ \\
 & P051436105309 & 59999.11 & 2453 & $615.1\pm{0.5}$ \\
 \hline
P0514361054 & P051436105401 & 59999.49 & 2157 & $618.2\pm{0.5}$ & $205.48\pm{0.29}$\\
 & P051436105402 & 59999.64 & 927 & $586.3\pm{0.8}$ \\
 & P051436105403 & 59999.78 & 965 & $577.5\pm{0.8}$ \\
 & P051436105404 & 59999.91 & 3155 & $588.7\pm{0.4}$ \\
 & P051436105405 & 60000.04 & 3074 & $675.2\pm{0.5}$ & $205.38\pm{0.21}$\\
 & P051436105406 & 60000.17 & 2621 & $592.4\pm{0.5}$ \\
 & P051436105408 & 60000.43 & 1512 & $592.8\pm{0.6}$ \\
 \hline
P0514361055 & P051436105502 & 60001.29 & 1293 & $575.9\pm{0.7}$ & $205.34\pm{0.25}$\\
 & P051436105504 & 60001.56 & 2170 & $538.8\pm{0.5}$ \\
 & P051436105505 & 60001.69 & 17 & $482.2\pm{5.4}$ \\
 & P051436105506 & 60001.82 & 2663 & $559.2\pm{0.5}$ & $205.20\pm{0.83}$\\
 & P051436105507 & 60001.95 & 1454 & $522.6\pm{0.6}$ \\
 & P051436105508 & 60002.09 & 2998 & $536.1\pm{0.4}$ \\
 \hline
P0514361056 & P051436105601 & 60003.99 & 3666 & $496.0\pm{0.4}$ & $205.26\pm{0.32}$\\
 & P051436105602 & 60004.13 & 2817 & $491.3\pm{0.4}$ \\
 & P051436105603 & 60004.27 & 1208 & $462.5\pm{0.6}$ \\
 & P051436105604 & 60004.40 & 3100 & $486.0\pm{0.4}$ \\
 & P051436105605 & 60004.53 & 1841 & $482.1\pm{0.5}$ & $205.18\pm{0.30}$\\
 & P051436105606 & 60004.66 & 822 & $526.1\pm{0.8}$ \\
 & P051436105607 & 60004.80 & 3422 & $458.7\pm{0.4}$ \\
 & P051436105608 & 60004.93 & 4760 & $447.4\pm{0.3}$ \\
 \hline
P0514361057 & P051436105701 & 60005.77 & 4116 & $428.1\pm{0.3}$ & $205.14\pm{0.39}$\\
 & P051436105702 & 60005.92 & 3540 & $435.9\pm{0.4}$ \\
 & P051436105703 & 60006.05 & 2965 & $455.7\pm{0.4}$ \\
 \hline
P0514361058 & P051436105801 & 60006.69 & 2894 & $415.2\pm{0.4}$ & $205.14\pm{0.42}$\\
 & P051436105802 & 60006.84 & 3362 & $393.6\pm{0.3}$ \\
 & P051436105803 & 60006.98 & 2843 & $435.1\pm{0.4}$ \\
 \hline
P0514361059 & P051436105901 & 60007.69 & 3131 & $355.4\pm{0.3}$ & $205.09\pm{0.60}$\\
 & P051436105902 & 60007.84 & 3374 & $406.6\pm{0.3}$ \\
 & P051436105903 & 60007.97 & 2793 & $412.9\pm{0.4}$ \\
 \hline
P0514361060 & P051436106001 & 60008.61 & 1200 & $404.4\pm{0.6}$ & $205.07\pm{0.38}$\\
 & P051436106002 & 60008.76 & 3203 & $364.2\pm{0.3}$ \\
 & P051436106003 & 60008.89 & 4452 & $354.4\pm{0.3}$ \\
 \hline
P0514361061 & P051436106101 & 60009.87 & 4262 & $329.4\pm{0.3}$ & $205.04\pm{0.48}$\\
 & P051436106102 & 60010.02 & 2695 & $344.8\pm{0.4}$ \\
 & P051436106103 & 60010.14 & 1239 & $305.3\pm{0.5}$ \\
 \hline
P0514361064 & P051436106401 & 60016.81 & 2507 & $229.7\pm{0.3}$ & $204.90\pm{0.90}$\\
 & P051436106402 & 60016.95 & 1684 & $197.1\pm{0.3}$ \\
 \hline
P0514361065 & P051436106501 & 60018.13 & 3364 & $195.0\pm{0.2}$ & $204.91\pm{0.26}$\\
 & P051436106502 & 60018.28 & 3262 & $199.2\pm{0.2}$ \\
 & P051436106503 & 60018.41 & 267 & $201.0\pm{0.9}$ \\
 \hline
P0514361066 & P051436106601 & 60019.99 & 716 & $198.5\pm{0.5}$ & $204.78\pm{0.64}$\\
 & P051436106602 & 60020.12 & 4287 & $171.3\pm{0.2}$ \\
 & P051436106603 & 60020.26 & 2276 & $167.3\pm{0.3}$ \\
 \hline
P0514361067 & P051436106701 & 60021.77 & 1592 & $152.8\pm{0.3}$ & $204.73\pm{1.05}$\\
 & P051436106702 & 60021.90 & 978 & $161.3\pm{0.4}$ \\
 & P051436106703 & 60022.04 & 1797 & $139.7\pm{0.3}$ \\
 \hline
P0514361068 & P051436106802 & 60023.10 & 6181 & $136.6\pm{0.1}$ & $204.85\pm{0.48}$\\
 \hline
P0514361071 & P051436107101 & 60024.55 & 2266 & $136.8\pm{0.2}$ & $204.72\pm{1.24}$\\
 & P051436107102 & 60024.69 & 1723 & $131.9\pm{0.3}$ \\
 & P051436107103 & 60024.82 & 947 & $137.0\pm{0.4}$ \\
 \hline
P0514361072 & P051436107201 & 60026.01 & 2848 & $118.4\pm{0.2}$ & $204.74\pm{0.78}$\\
 & P051436107202 & 60026.15 & 2242 & $117.9\pm{0.2}$ \\
 & P051436107203 & 60026.28 & 1206 & $104.6\pm{0.3}$ \\
 \hline
P0514361073 & P051436107301 & 60027.79 & 775 & $108.2\pm{0.4}$ & $204.73\pm{1.27}$\\
 & P051436107302 & 60027.93 & 1114 & $102.7\pm{0.3}$ \\
 & P051436107303 & 60028.06 & 3513 & $103.6\pm{0.2}$ \\
 \hline
 P0514361074 & P051436107401 & 60029.25 & 1028 & $86.1\pm{0.3}$ & $204.75\pm{1.69}$\\
 & P051436107402 & 60029.39 & 2028 & $97.7\pm{0.2}$ \\
 & P051436107403 & 60029.52 & 2434 & $104.2\pm{0.2}$ \\

    %\tablecomments{The QPOs analyzed in this paper are highlighted in bold text.}
    %\footnote{The QPOs analyzed in this paper are highlighted in bold text.}
\end{longtable}
\end{center}

\begin{center}
\onecolumn
\setlength{\tabcolsep}{5mm}
\begin{longtable}{ccccc}

    \caption{\textit{NICER} observations log used in this paper. The count rates in the 2$-$10\,keV energy range and the periods are listed. }
    
    \label{tab:nicer}\\
    
    \hline
    ObsIDs & Start time (MJD) & Exposure time (s)& Count Rate (cts $\rm{s^{-1}}$) & Period (s)\\
    \hline   
    \endfirsthead
    \hline
    ObsIDs & Start time (MJD) & Exposure time (s)&Count Rate (cts $\rm{s^{-1}}$)  & Period (s) \\
    \hline   
    \endhead
    \hline 
    \multicolumn{4}{r}{to be continued}\\
\hline
    \endfoot
    \hline
    \endlastfoot
    
5203610101 & 59942.79 & 3786 & $173.8\pm{0.2} $ & $208.03\pm{0.97}$\\
 5203610102 & 59942.99 & 8256 & $201.4\pm{0.2} $ & $208.02\pm{0.18}$\\
 5203610103 & 59944.15 & 6535 & $214.9\pm{0.2} $ & $207.99\pm{0.39}$\\
 5203610104 & 59945.44 & 1985 & $242.2\pm{0.3} $ & $208.02\pm{0.36}$\\
 5203610105 & 59946.03 & 5339 & $255.5\pm{0.2} $ & $207.96\pm{0.17}$ \\
 5203610106 & 59947.65 & 2697 & $258.4\pm{0.3} $ & $207.90\pm{1.73}$\\
 5203610107 & 59948.04 & 2380 & $188.7\pm{0.3} $& $207.91\pm{0.19}$\\
 5203610108 & 59949.65 & 5525 & $226.0\pm{0.2} $ & $207.86\pm{0.47}$\\
 5203610109 & 59950.02 & 15481 & $213.4\pm{0.1} $ & $207.87\pm{0.18}$\\
 5203610110 & 59950.99 & 15256 & $198.5\pm{0.1} $ & $207.86\pm{0.23}$\\
 5203610111 & 59952.02 & 7770 & $161.5\pm{0.1} $ & $207.84\pm{0.20}$\\
 5203610112 & 59953.11 & 4771 & $160.0\pm{0.2} $ & $207.84\pm{0.22}$\\
 5203610113 & 59954.16 & 3685 & $146.5\pm{0.2} $& $207.81\pm{0.19}$\\
 5203610114 & 59955.05 & 3745 & $125.2\pm{0.2} $ & $207.81\pm{0.41}$  \\
 5203610115 & 59956.02 & 2762 & $105.7\pm{0.2} $ & $207.85\pm{0.25}$\\
 5203610116 & 59957.11 & 3601 & $97.2\pm{0.2} $ & $207.78\pm{0.39}$\\
 5203610117 & 59958.08 & 2590 & $102.9\pm{0.2} $&~.~.~.\\
 5203610118 & 59959.18 & 1746 & $157.5\pm{0.3} $&~.~.~.\\
 5203610119 & 59960.34 & 1511 & $184.3\pm{0.3} $&~.~.~.\\
 5203610120 & 59961.44 & 1914 & $236.3\pm{0.4} $&~.~.~.\\
 5203610121 & 59962.21 & 1773 & $274.6\pm{0.4} $&~.~.~.\\
 5203610122 & 59963.37 & 1429 & $324.4\pm{0.5} $&~.~.~.\\
 5203610123 & 59964.02 & 1323 & $345.0\pm{0.5} $&~.~.~.\\
 5203610124 & 59965.25 & 3548 & $385.8\pm{0.3} $& $207.60\pm{0.22}$\\
 5203610125 & 59966.02 & 3455 & $408.6\pm{0.3} $ & $207.58\pm{0.16}$\\
 5203610126 & 59967.05 & 1766 & $413.7\pm{0.5} $ & $207.53\pm{4.21}$\\
 5203610127 & 59968.15 & 5312 & $449.4\pm{0.3} $ & $207.52\pm{0.22}$\\
 5203610128 & 59969.11 & 4677 & $439.7\pm{0.3} $ & $207.46\pm{0.27}$\\
 5203610129 & 59970.08 & 4556 & $495.7\pm{0.3} $ & $207.40\pm{0.15}$\\
 5203610130 & 59971.18 & 5417 & $535.5\pm{0.3} $ & $207.35\pm{0.21}$\\
 5203610131 & 59972.01 & 4454 & $650.2\pm{0.4} $ & $207.29\pm{0.15}$\\
 5203610132 & 59973.05 & 4060 & $721.7\pm{0.4} $ & $207.21\pm{0.17}$\\
 5203610133 & 59974.14 & 5354 & $757.2\pm{0.4} $ & $207.09\pm{0.19}$\\
 5203610134 & 59975.11 & 3507 & $763.1\pm{0.5} $ & $206.93\pm{0.20}$\\
 5203610135 & 59976.01 & 6989 & $805.9\pm{0.3} $ & $206.90\pm{0.15}$\\
 5203610136 & 59977.05 & 5099 & $854.8\pm{0.4} $ & $206.79\pm{0.15}$\\
 5203610137 & 59978.01 & 11959 & $799.3\pm{0.3} $ & $206.73\pm{0.26}$\\
 5203610138 & 59979.17 & 5629 & $822.6\pm{0.4} $ & $206.62\pm{0.24}$\\
 5203610139 & 59980.01 & 6194 & $749.7\pm{0.3} $& $206.51\pm{0.20}$\\
 5203610140 & 59981.11 & 7441 & $736.9\pm{0.3} $& $206.44\pm{0.26}$\\
 5203610141 & 59982.01 & 11313 & $715.4\pm{0.3} $& $206.37\pm{0.16}$\\
 5203610142 & 59983.05 & 7778 & $685.5\pm{0.3} $ & $206.30\pm{0.19}$\\
 5203610143 & 59984.08 & 11269 & $660.9\pm{0.2} $& $206.24\pm{0.23}$\\
 5203610144 & 59985.50 & 4309 & $636.8\pm{0.4} $& $206.14\pm{0.29}$\\
 5203610145 & 59986.02 & 11726 & $638.9\pm{0.2} $& $206.11\pm{0.17}$\\
 5203610146 & 59987.24 & 2758 & $611.8\pm{0.5} $ & $206.05\pm{0.29}$\\
 5203610147 & 59988.02 & 3868 & $631.0\pm{0.4} $ & $206.00\pm{0.15}$\\
 5203610148 & 59989.00 & 6620 & $611.5\pm{0.3} $ & $205.95\pm{0.20}$\\
 5203610149 & 59990.02 & 15075 & $594.5\pm{0.2} $& $205.89\pm{0.16}$\\
 5203610150 & 59991.11 & 9389 & $562.5\pm{0.2} $& $205.84\pm{0.22}$\\
 5203610151 & 59992.22 & 2463 & $569.0\pm{0.5} $& $205.77\pm{0.25}$\\
 5203610152 & 59993.19 & 4163 & $572.2\pm{0.4} $& $205.73\pm{0.23}$\\
 5203610153 & 59994.01 & 6335 & $553.5\pm{0.3} $ & $205.69\pm{0.17}$\\
 5203610154 & 59995.04 & 9434 & $546.1\pm{0.2} $& $205.64\pm{0.23}$\\
 5203610158 & 60001.29 & 2181 & $559.5\pm{0.5} $ & $205.31\pm{0.51}$\\
 6203610101 & 60005.66 & 1605 & $407.8\pm{0.5} $& $205.22\pm{1.15}$\\
 6203610102 & 60006.44 & 1088 & $450.9\pm{0.6} $&~.~.~.\\
 6203610103 & 60007.03 & 5405 & $411.2\pm{0.3} $ & $205.11\pm{0.16}$\\
 6203610104 & 60008.37 & 3789 & $405.3\pm{0.3} $& $205.07\pm{0.35}$\\
 6203610105 & 60009.21 & 6983 & $390.5\pm{0.2} $& $205.06\pm{0.33}$\\
 6203610106 & 60010.31 & 2590 & $369.2\pm{0.4} $ & $205.03\pm{0.32}$\\
 6203610107 & 60011.21 & 6722 & $346.0\pm{0.2} $& $204.99\pm{0.20}$\\
 6203610108 & 60012.24 & 5292 & $334.4\pm{0.3} $& $205.01\pm{0.47}$\\
 6203610109 & 60013.03 & 3094 & $324.2\pm{0.3} $& $204.96\pm{0.24}$\\
 6203610110 & 60014.06 & 4442 & $314.6\pm{0.3} $& $204.93\pm{0.25}$\\
 6203610111 & 60015.02 & 3190 & $309.1\pm{0.3} $& $204.91\pm{0.26}$\\
 6203610112 & 60016.12 & 3013 & $284.8\pm{0.3} $& $204.89\pm{0.22}$\\
 6203610113 & 60017.21 & 2999 & $266.6\pm{0.3} $& $204.90\pm{0.28}$\\
 6203610114 & 60018.18 & 3628 & $274.3\pm{0.3} $& $204.87\pm{0.26}$\\
 6203610115 & 60019.20 & 7252 & $237.3\pm{0.2} $& $204.87\pm{0.63}$\\
 6203610116 & 60020.04 & 13922 & $234.0\pm{0.1} $& $204.84\pm{0.16}$\\
 6203610117 & 60021.08 & 8270 & $224.0\pm{0.2} $& $204.83\pm{0.17}$\\
 6203610118 & 60021.98 & 7739 & $202.4\pm{0.2} $& $204.82\pm{0.16}$\\
 6203610119 & 60023.15 & 8534 & $199.0\pm{0.2} $& $204.81\pm{0.15}$\\
 6203610120 & 60023.98 & 7635 & $204.8\pm{0.2} $& $204.81\pm{0.11}$\\
 6203610121 & 60025.10 & 10040 & $192.3\pm{0.1} $& $204.79\pm{0.25}$\\
 6203610122 & 60025.99 & 6237 & $203.7\pm{0.2} $& $204.79\pm{0.14}$\\
 6203610123 & 60027.01 & 7786 & $147.6\pm{0.1} $& $204.78\pm{0.11}$\\
 6203610124 & 60028.00 & 7271 & $167.4\pm{0.2}$ & $204.78\pm{0.18}$\\
 6203610125 & 60029.02 & 7916 & $162.9\pm{0.1}$ & $204.78\pm{0.15}$\\
 6203610126 & 60030.05 & 3230 & $155.0\pm{0.2} $& $204.80\pm{0.23}$\\
 6203610127 & 60031.08 & 4044 & $123.1\pm{0.2}$ & $204.78\pm{0.20}$\\

    %\tablecomments{The QPOs analyzed in this paper are highlighted in bold text.}
    %\footnote{The QPOs analyzed in this paper are highlighted in bold text.}
\end{longtable}
\end{center}

%%%%%%%%%%%%%%%%%%%%%%%%%%%%%%%%%%%%%%%%%%%%%%%%%%

% Don't change these lines
\bsp	% typesetting comment
\label{lastpage}
\end{document}